%% file: main.tex
\documentclass[sigconf]{acmart}
\AtBeginDocument{%
  }

\copyrightyear{2026}
\acmYear{2026}
\setcopyright{cc}
\acmConference[CCS '26]{Proceedings of the 2026 ACM SIGSAC Conference on Computer and Communications Security}{November 15--19, 2026}{The Hague, Netherlands}
\acmBooktitle{Proceedings of the 2026 ACM SIGSAC Conference on Computer and Communications Security (CCS '26), November 15--19, 2026, The Hague, Netherlands}
\acmDOI{10.1145/3830454.3846653}
\acmISBN{979-8-4007-2871-6/2026/11}
\usepackage{url}
\usepackage{balance} 
\usepackage{booktabs}
\usepackage{tabularx}
\usepackage[ruled,vlined]{algorithm2e}
\usepackage{listings}
\usepackage{comment}
\usepackage{enumitem}
\usepackage{multirow}
\usepackage{tcolorbox}
\tcbuselibrary{theorems}
\usepackage{float}
\usepackage{graphicx}

\newcounter{finding}
\renewcommand{\thefinding}{\arabic{finding}}
\newtcolorbox{findingbox}{
  colback=gray!10!white,
  colframe=gray!50!black,
  boxrule=0.5pt,
  before upper=\refstepcounter{finding}\textbf{Finding~\thefinding. }\ignorespaces
}

\newcommand{\framework}{IoTAbandonGuard\xspace}
\newcommand{\parhead}[1]{\medskip\Parhead{#1}}
\newcommand{\Parhead}[1]{\noindent\textbf{#1}\hskip 0.5em\relax}

\begin{document}

\title{When Apps Outlive Vendors: Security Implications of IoT Abandonware}


\author{Dayeon Kang}
\affiliation{%
  \institution{University of Massachusetts Amherst}
  \city{Amherst, MA}
  \country{USA}}
\email{dayeonkang@umass.edu}

\author{Elvis Yeboah-Duako}
\affiliation{%
  \institution{University of Massachusetts Amherst}
  \city{Amherst, MA}
  \country{USA}
}
\email{eyeboahduako@umass.edu}

\author{Sachin Thomas}
\affiliation{%
  \institution{University of Massachusetts Amherst}
  \city{Amherst, MA}
  \country{USA}}
\email{sachinthomas@umass.edu}

\author{Pubali Datta}
\affiliation{%
  \institution{University of Massachusetts Amherst}
  \city{Amherst, MA}
  \country{USA}}
\email{pdatta@umass.edu}

\renewcommand{\shortauthors}{Kang et al.}

\begin{abstract} 
As the Internet of Things (IoT) market continues to expand, many companion apps are being published in app stores, raising security concerns for those whose vendors have abandoned support. Even after vendors discontinue support, such applications frequently remain operational on users' mobile devices, continue to interface with users' IoT devices and collect user data without receiving security updates.  
This leaves known and newly discovered vulnerabilities unmitigated, increasing risks of remote exploitation, unauthorized device access, and prolonged data misuse. We define these abandoned applications as "IoT abandonware" and present the first large-scale measurement study of the security risks associated with discontinued applications.

We analyze 61,500 IoT companion Android applications that had not been updated for at least two years or were no longer in service as of March 2025. From decompiled binaries, we extracted latent and embedded resources (e.g., bundled libraries, domain names, and permissions), and assessed their security implications. First, we identify outdated dependencies with post-abandonment CVE reports and discover domains vulnerable to takeover or data exfiltration. Second, 
we perform static data-flow analysis to trace how sensitive data, inferred from the extracted permissions, propagates to broken or hijackable external endpoints. 
We found that persistent analytics and third-party trackers continue aggregating user data and device telemetry long after vendor control lapses, creating data flows that adversaries can redirect or abuse. Overall, we identified security risks in 73.6\% of our dataset, with 30 of the top 1,000 most-installed apps sending data to broken external endpoints.
\end{abstract}



\maketitle

\input{sections/introduction}
\input{sections/background}

\input{sections/motivation}
\input{sections/methodology}

\input{sections/eval}

\input{sections/limitations}
\input{sections/relatedwork.tex}

\input{sections/conclusion}

\bibliographystyle{ACM-Reference-Format}
\bibliography{main}

\appendix


\section{Ethical Considerations}
This research adheres to established ethical standards for security research involving real-world applications. 
We follow the previous studies’ method of real-world application collection and storing format~\cite{nan2023you, jin2022understanding, neupane2022data, schmidt2023iotflow} while handling the abandoned Android IoT application dataset.
We collected publicly available APK files and associated metadata from AndroZoo~\cite{AndroZoo}. All applications analyzed in this study were already publicly accessible, and no proprietary or restricted data sources were utilized.

The principal stakeholders in this study are: (1) end users of IoT companion applications, who may experience security or privacy harms if vulnerabilities are exploited; (2) IoT device and app vendors, whose products are analyzed and whose reputations or operations could be affected by disclosure. 
Considering ethical principles, all collected datasets and security analysis outputs remain under our control, and we did not exploit any security vulnerabilities and attempt attacks against the collected Android IoT abandonware.
As we study application-level vulnerabilities, we do not collect, possess, or process any information involving user data or user behavior patterns. To mitigate the impact on vendors, we do not report the exact name of IoT companion apps and their vulnerabilities revealed via our security analysis. 

\subsection{Responsible Disclosure}

We made responsible disclosures to the Google App Security Improvement program~\cite{SecurityImprovement}, which is a service provided to Google Play app developers, by sending emails to \url{security+asi@android.com} to report potential security issues in IoT applications revealed from our analysis on April 30, 2026. The team informed us that they only accept vulnerability reports for Google's own products and could not assist with third-party app disclosure. We therefore contacted IoT app vendors directly via contact emails listed on AndroZoo, starting on June 15.

From AndroZoo, we identified 27,474 app--including removed, unmaintained (over two years), and active apps--with associated developer email addresses. These apps are expected to be highly impacted by vulnerabilities, which are HIGH/CRITICAL CVEs or malicious/owner-changed endpoints. During manual verification with vendors, we found CVE-2012-0754 (Adobe Flash Player) was a false positive introduced by APK decompilation artifacts, which we excluded. We then emailed 14,635 apps from June 16th: 12,899 for removed apps, 1,597 for apps unmaintained for 2+ years, and 139 for active apps.

As of August 15, 2,576 emails (18\%) had bounced due to invalid addresses. We received 135 (1\%) replies: 9 vendors removed their apps from Google Play, 5 confirmed the reported CVEs were not applicable to their apps, and 11 vendors acknowledged the issue and are actively working on fixes. Many of the remaining respondents requested further analysis details, and we are providing them. 

As one example, our report of CVE-2025-56400 and CVE-2025-56557 (Tuya SDK vulnerabilities) led one vendor of a discontinued app in the Google Play Store to confirm it. The vendor responded that the issue is server-side and cannot be fixed client-side because they cannot fix the bundled Tuya SDK directly, but the issue was resolved since Tuya has remediated it on their cloud platform.

\section{Appendix}

\subsection{Outdated Dependencies Analysis} \label{sec:appendix-cve-alg}
Algorithm~\ref{alg:cve-search} details the CVE searching procedure used to identify vulnerable dependencies, as described in Section~\ref{sec:cve-analysis}.

\begin{algorithm}[t]
    \caption{CVE Searching}
    \label{alg:cve-search}
    \KwIn{Decompiled Java code directory name \texttt{dir}}
    \KwIn{META files of libraries $M = \{m_{lib_1}, m_{lib_2}, ..., m_{lib_m}\}$}
    \KwIn{$app$'s last updated date list $U = \{u_1, u_2, ..., u_{n}\}$}
    \KwOut{Found CVE \texttt{record} or \texttt{NULL}}

    \SetKwFunction{FGetCVE}{get\_cve}
    \SetKwProg{Fn}{Function}{:}{}
    \Fn{\FGetCVE{dir}}{
        $found\_cves \gets$ NVDSearch($dir$)\; 
        \If{len($found\_cves$) $>$ threshold //we used 50}{
            $dir \gets$ GetSubDir($dir$)\;
            $record \gets$ \FGetCVE{dir}\;
        }
        \Else{
            \If{$dir \in found\_cves.cpe$}{
                $cve\_ver \gets found\_cves.cpe$\;
            } \Else{$cve\_ver \gets found\_cves.published$\;}
            \If{$dir \in \{lib_1, lib_2, ..., lib_m\}$}{
                    $lib\_ver \gets m_{lib_{dir}}$\;
                }\Else{$lib\_ver \gets U[app]$\;}
            \If{$cve\_ver > lib\_ver$ //$cve\_ver$ is recent}{
                \Return{$found\_cves$}
            }
        }
    }
\end{algorithm}

\subsection{Taxonomy of Data-Flow Severity}~\label{sec:appendix-taxonomy}
Table~\ref{tab:data-type-taxonomy} presents the four-tier severity 
classification used to assess data-collecting API methods in our data flow analysis.

\begin{table}[]
    \centering
    \caption{Four-tier risk classification of data-collecting API methods.}
    \label{tab:data-type-taxonomy}
    \begin{tabularx}{\columnwidth}{@{}lX@{}}
        \toprule
        Severity & API method(s) \\ \midrule
        Critical &
        \texttt{getSubscriberId}, \texttt{getDeviceId}, \texttt{getSimSerialNumber},
        \texttt{getLine1Number}, \texttt{getLastKnownLocation}, \texttt{getLatitude},
        \texttt{getLongitude}, \texttt{getAddress}, \texttt{getMacAddress},
        \texttt{getSSID} \\

        High &
        \texttt{getInstalledApplications}, \texttt{getInstalledPackages},
        \texttt{queryIntentActivities}, \texttt{queryIntentServices},
        \texttt{queryContentProviders}, \texttt{queryBroadcastReceivers},
        \texttt{getCountry}, \texttt{getTimeZone} \\

        Medium &
        \texttt{getCid}, \texttt{getLac} \\

        Low &
        \texttt{getInputStream()}, \texttt{read}, \texttt{execute()},
        \texttt{getCharacteristics()}, \texttt{getCharacteristic(java.util.UUID)},
        \texttt{source\_function} \\
        \bottomrule
    \end{tabularx}
\end{table}

\subsection{Observations on Cryptographic Practice} \label{appendix:crypto}
\begin{figure}[h]
  \centering
  \includegraphics[width=\linewidth]{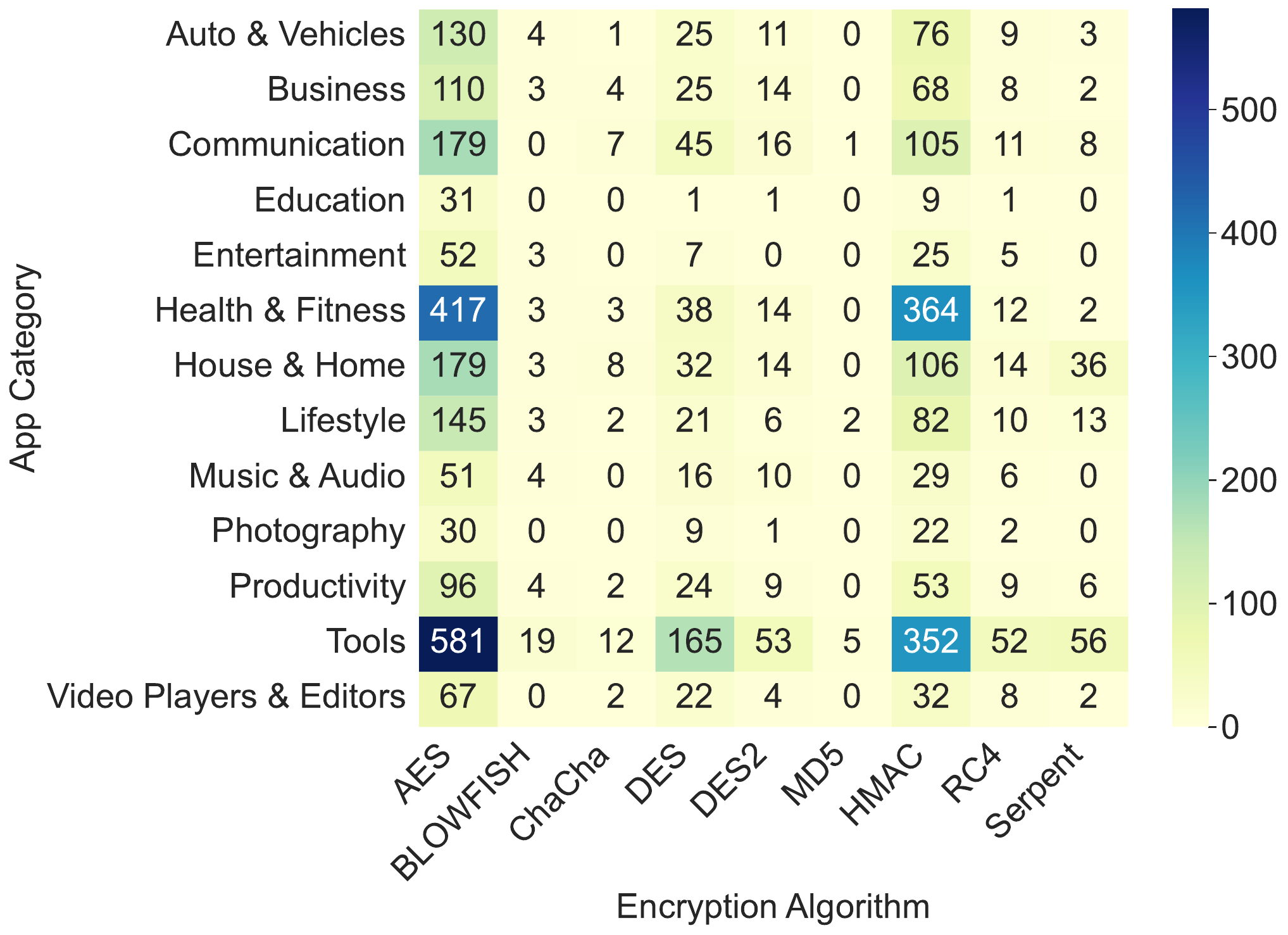}
  \caption{Dispersion of different encryption algorithms across apps not updated for more than 2 years.}
  \label{fig:encrypt-alg-appendix}
\end{figure}

Figure~\ref{fig:encrypt-alg-appendix} shows the distribution of encryption 
algorithms across apps unmaintained for over two years, complementing the 
cryptographic analysis in Section~\ref{sec:crypto}.

\end{document}

%% file: sections/introduction.tex
\section{Introduction}

Mobile companion applications serve as the principal user interface for most consumer Internet of Things (IoT) devices. The rapid expansion of the IoT market has, in turn, driven a sharp rise in the number of such applications being released on public marketplaces. Concretely, the number of connected IoT devices has grown from 18.8 billion at the end of 2024 \cite{iotbusinessnewsState2024} to 21.1 billion\cite{iotanalyticsNumberConnected} in 2025, reflecting the pace and scale of this proliferation. This competitive pressure has led many developers to prioritize time-to-market over sustained maintenance, leaving a significant share of companion applications insufficiently patched and susceptible to known vulnerabilities. 

On the Google Play Store alone, 5\% of the top 1,000 mobile applications are updated less than once per year~\cite{42mattersUpdateFrequency}, and a recent study found that 869,000 apps on Google Play Store had not received any update in over two years~\cite{appleinsiderDevelopersAbandoning}. This degree of abandonment is particularly alarming given that users expect that IoT devices remain in active use for roughly a decade on average~\cite{iot-longevity}, a lifespan that far outlasts the maintenance commitments from vendors for companion applications.
Wang et al. \cite{Wang_Jiang_Chang_Zhou_Hou_Luo_Wu_Ren_2021} reported more than 2 million end-of-life devices still in active deployment, with nearly 300,000 remaining operational even five years after their companion applications were abandoned. Consequently, many users are left relying on unmaintained or delisted applications to manage sensitive IoT data, significantly broadening their security exposure. This concern is reflected in user sentiment, where 58\% of respondents express concern about IoT app security compared to 53\% for general mobile apps \cite{iottechnewsStudyReveals} -- a gap driven, in large part, by the sensitive nature of data that IoT companion apps routinely collect, including feeds from webcams, medical sensors, smart-home devices, and GPS trackers.

In this work, we define IoT-abandonware as IoT companion apps that are not available in the app store (as of March 2025) or have not been updated for more than two years (last update in March 2023). Prior work has examined related phenomena, such as studying abandoned Internet resources (i.e., domains and IP addresses) used by Android mobile apps ~\cite{pariwono2018don} and discovering known vulnerabilities (e.g., Common Vulnerabilities and Exposures (CVE) and Exploit Database (EDB)) in end-of-Life embedded devices~\cite{10321684, Wang_Jiang_Chang_Zhou_Hou_Luo_Wu_Ren_2021}. However, there has been no comprehensive study of abandoned companion apps that remain installed on users’ devices.
To fill this gap, we curated a dataset of 61,500 abandoned IoT apps and conducted a large‑scale, holistic security and privacy measurement of those apps. Our study is driven by the following research questions. 




\begin{itemize}[leftmargin=4mm]
    \item \textbf{RQ1}: How widespread are abandoned IoT app installs and use?
    \item \textbf{RQ2}: How prevalent are vulnerabilities in abandoned IoT apps
that stem from artifacts embedded in unmaintained apps?
    \item \textbf{RQ3}: What is the risk of sensitive data exposure from abandoned IoT apps on users’ devices?
    \item \textbf{RQ4}: Are the security issues in abandoned IoT apps unique to abandonment?
\end{itemize}

To answer the above questions, we first built the IoT abandonware dataset by collecting mobile apps from AndroZoo \cite{allix2016androzoo}, and classifying them using the BERT-based mobile-IoT app classifier proposed by Jin et al. as part of their IoTSpotter framework~\cite{jin2022understanding}. This classifier identifies mobile apps that are used as companions or automation providers to IoT devices. Next, we filtered out apps that fit our definition of IoT-abandonware to build our dataset.

In our analysis of 61,500 abandoned IoT applications, we focus on security risks arising from outdated and CVE-reported dependencies, at-risk domains, and broken data flows.
We assess both internal app properties and external ecosystem factors that jointly determine an abandoned companion app’s security posture. In our analysis pipeline, we extract app metadata from marketplaces (e.g., download counts and user reviews), static manifest information (permissions and declared sensor accesses), and program structure from decompiled apps. The information from marketplaces on downloads denotes the prevalence of IoT abandonware. We conduct two complementary analyses with decompiled apps. First, we identify latent, embedded resources within the app (e.g., bundled libraries and hard‑coded domain names) and extract the static sinks and sources referenced by the app’s code. Second, we apply static data‑flow analysis to trace how sensitive inputs (sensor values, camera/microphone streams, GPS, credentials) propagate through program paths to external sinks or to locally persisted artifacts. 


A substantial body of prior work has investigated data flow analysis in Android applications through both dynamic and static analysis techniques~\cite{schmidt2023iotflow, enck2014taintdroid, Arzt2014FlowDroidPC, wang2019looking,nan2023you}. Our flow analysis approach is grounded in IoTFlow~\cite{schmidt2023iotflow}, a static analysis framework tailored for IoT applications.
We also analyze protections applied to network communications originating from sensitive data flows, inspecting protocol usage and the presence and quality of cryptographic primitives implemented in apps. When a sensitive data flow terminates at an external sink over an inadequately protected channel (e.g., plaintext HTTP or deprecated cryptographic algorithms), the data is susceptible to interception, tampering, or unauthorized access.

Building on this analysis pipeline, we systematically assess the residual security risks introduced by IoT application abandonment across three dimensions. First, we enumerate bundled and referenced third-party libraries, cross-reference them with CVE databases, and flag unpatched dependencies that expand the attack surface following the application's final release. Second, we evaluate the reachability and integrity of hard-coded third-party endpoints embedded in abandoned APKs, examining domain ownership transitions and consulting blocklists and reputation feeds. We define domains that are unreachable, owner-changed, or blocklisted as at-risk domains. Endpoints that become unreachable or are repurposed by new registrants represent a pathway for data exfiltration and adversarial impersonation. Third, we identify sensitive input flows directed at broken or compromised backend infrastructure, exposing data to leakage, unauthorized access, and potential takeover.

In our dataset of 61,500 abandoned IoT companion apps, we found that 73.6\% rely on CVE-reported libraries, every app has at least one at-risk domain, and 38.4\% of the data flow reaches broken sinks or sources, transferring sensitive data. Overall, 73.6\% of these abandoned apps face documented security risks, and the top 1,000 installed apps have slightly higher risks than the overall trend. Our contributions are as follows:
\begin{itemize}[leftmargin=4mm]
    \item We perform the first large-scale empirical study of abandoned IoT companion applications. 
    \item We curate 61,500 abandoned Android IoT apps and analyze their security risks. We release the measurement code and dataset~\footnote{The artifact is available at: \url{https://bitbucket.org/umass-lab/iot-abandonware-analysis}}.
    \item Using our analysis pipeline, we find that 73.6\% of apps in our dataset contain CVE-reported dependencies, at-risk domains, or data flows to broken backends over weak network channels.
\end{itemize}

%% file: sections/background.tex
\section{Backgroud \& Motivation} 
\subsection{IoT Abandonment Factors and Risks}
\label{sec:background}

The abandonment of IoT apps is intrinsically linked to the lifecycle of their hardware counterparts. Factors such as commercial failures, strategic decisions (e.g., releasing a newer model and sunsetting older infrastructure), or temporary service disruption often lead to "abandonware"--applications that remain functional but are no longer maintained by the vendor. Handling discontinuation varies: while some vendors provide end-of-life (EoL) notices and final updates, others offer trade-in programs or simply take no action. 
Since IoT devices are costly or difficult to replace, users may take actions to perpetuate outdated apps: keep devices and their companion apps long after official support ends~\cite{iot-longevity}, use local functionality without cloud dependencies and awareness of backbone abandonment, and take collective actions and extend the app longevity for shared household or multi-user devices. Some users, on the other hand, may be totally unaware of the discontinuation or abandonment of the IoT app or product.

The technical risks of abandonware stem from variable and wide attack surfaces. Modern Android apps rely on a complex interplay of system permissions (e.g., sensor and storage access), third-party libraries for telemetry and ads, and network endpoints for API calls and cloud communication. Without active maintenance, these components accumulate unpatched vulnerabilities, providing entry points for exploitation~\cite{chatzoglou2022iotapps, 208133}.

%% file: sections/motivation.tex
\subsection{Motivation}
\label{sec:motivation}
Consider an abandoned smart-home application\footnote{App name undisclosed 
for ethical reasons} from our dataset (Sec.\ref{sec:dataset}) that controls 
domotics, Wi-Fi-connected smart plugs, and household appliances. In addition 
to remote control, it provides simultaneous management of multiple devices, 
timer-based automation, and one-tap device sharing among family members, 
making it a convenient hub for home automation users. The first release was 
in April 2019, and it accumulated over 5,000 installs before receiving its 
final update in April 2020, after which development went silent. The app was 
never formally discontinued or removed from app stores; it simply stopped 
being maintained. Despite this, it remains installed on thousands of user 
devices, and the smart home appliances it manages continue to communicate 
with hard-coded cloud endpoints, operating as though nothing has changed.

\subsubsection{Risks of Using Outdated Libraries}
Static analysis of the app's APK reveals a dependency on a vulnerable version 
of the \textit{FasterXML Jackson Databind} library (v2.9.1), a widely used 
Java JSON serialization library. Because the app enables default typing in its 
\texttt{ObjectMapper} configuration, it is susceptible to unsafe 
deserialization attacks. Two representative critical vulnerabilities 
in this version are \textit{CVE-2017-17485} (CVSS~9.8, 
Critical)~\cite{nistCVE201717485}, which allows unauthenticated remote code 
execution via a Spring gadget chain (affecting versions 2.9.x through 2.9.3), and 
\textit{CVE-2019-14379} (CVSS~9.8, Critical)~\cite{nistCVE201914379}, which 
enables remote code execution when \textit{ehcache} is present on the classpath (affecting versions prior to 2.9.9.2)\footnote{See related 
issue: \url{https://github.com/FasterXML/jackson-databind/issues/2387}}. Both 
vulnerabilities are triggered by submitting a crafted JSON payload that 
abuses the enabled default typing during deserialization. Although patches 
have been available since July 2019, the app has not been updated since 
April 2020 and therefore remains unpatched and exposed.

\Parhead{Attack Scenario.}
We find that the app implements a Tuya-based command-and-control protocol for 
device management and subscribes to device command topics at a hard-coded 
broker endpoint. An attacker who gains access to the broker (e.g., via 
credential stuffing against reused credentials) can publish a malicious JSON 
payload to a device command topic:

\begin{lstlisting}
{
  "command": "update_config",
  "timestamp": 1699824000,
  "payload": {
    "@class": "com.sun.org.apache.xalan.internal
               .lib.ExploitUtils",
    "method": "executeCommand",
    "args": ["sh", "-c", 
             "curl http://attacker.com/payload.sh | sh"]}
}
\end{lstlisting}

\Parhead{Security Breach Impact.}
After deserializing the payload, the gadget chain executes arbitrary commands 
with the app's declared permissions (\texttt{ACCESS\_FINE\_LOCATION}, 
\texttt{READ\_EXTERNAL\_STORAGE}, \texttt{READ\_CONTACTS}, and 
\texttt{CHANGE\_WIFI\_STATE}). This enables the attacker to exfiltrate stored 
credentials, enumerate the user's home network, and pivot to other smart 
devices on the same local segment. Had the app been maintained and updated 
from the vulnerable Jackson Databind release to a patched version (v2.9.9.2+, 
available since July 2019), this attack surface would have been closed. 
Nevertheless, the abandoned app remains indefinitely exposed.

\subsubsection{Risks of Domain Takeover}
Suppose a threat actor decompiled the app's APK and identified hard-coded 
endpoints such as \textit{\small 
\url{"https://api.app-domain.com/v1/devices/telemetry"}} and \textit{\small 
\url{"https://analytics.app-domain.net/events/batch"}} in the app. 
Additionally, upon inspecting the domains \texttt{app-domain.com} and 
\texttt{app-domain.net}, the attacker finds that they have expired following 
the vendor's abandonment of the product, and the application implements no 
certificate pinning for these endpoints, relying solely on standard Android 
system certificate validation. The attacker registers the expired domains and 
obtains valid TLS certificates via Let's Encrypt.

\Parhead{Attack Scenario.}
The original app's implementation transmits device telemetry on a periodic 
schedule, including:
\begin{lstlisting}
{
    "device_id":   "DV-A3F21C",
    "user_id":     "USR04872",
    "timestamp":   1699824000,
    "devices": [
        {"name": "Living Room Plug",
         "state": "ON", "power_w": 142.3},
         {"name": "Front Door Lock",
         "state": "LOCKED"}
        ...
    ],
    "schedules": [
        {"device": "Living Room Plug",
         "on":  "07:30", "off": "23:00"}
    ],
    "location":   [41.9028, 12.4964],
    "shared_with": [
        {"name":     "Marco Rossi",
         "relation": "spouse",
         "email":    "m.rossi@example.com"}
    ]
}
\end{lstlisting}
By reconstructing the API contract through reverse engineering, the attacker 
hosts a malicious server that mimics the original backend, silently logs all 
inbound data, and returns a valid HTTP~200 response, which the app accepts 
and continues normal operation, unaware that it is now streaming to an 
attacker-controlled server.

\Parhead{Privacy Breach Impact.}
Over time, this silent redirection enables the continuous exfiltration of 
fine-grained home-occupancy patterns: device-by-device power-consumption logs 
reveal daily routines, sleep schedules, and periods of absence, while 
shared-user records expose household members' identities and contact details. 
Real-time device state (e.g., whether door locks are engaged) further creates 
direct physical security risks. This scenario demonstrates that even a modest 
smart-plug controller, when abandoned or unmaintained, can become a 
long-running source of behavioral surveillance, enabling burglary planning, 
identity theft, and targeted social engineering against all household members.

\subsubsection{Take-away.}
These two scenarios (arbitrary code execution via an unpatched 
dependency and continuous data exfiltration via domain takeover) are 
not artifacts of a single poorly implemented app. They are structural 
consequences of abandonment: once a developer ceases maintenance, 
vulnerabilities accumulate, infrastructure lapses, and no mechanism 
exists to warn or protect the users still running the app. This raises 
a broader empirical question: \textit{how widespread are such risks 
across the population of abandoned smart-home apps, and what classes of 
vulnerability dominate?} The remainder of this paper presents a 
systematic study designed to answer these questions at scale.

%% file: sections/methodology.tex
\lstset{
  basicstyle=\ttfamily\small,
  breaklines=true,
  breakatwhitespace=true,
  columns=fullflexible,
}

\begin{figure}
    \centering
    \includegraphics[width=0.8\linewidth]{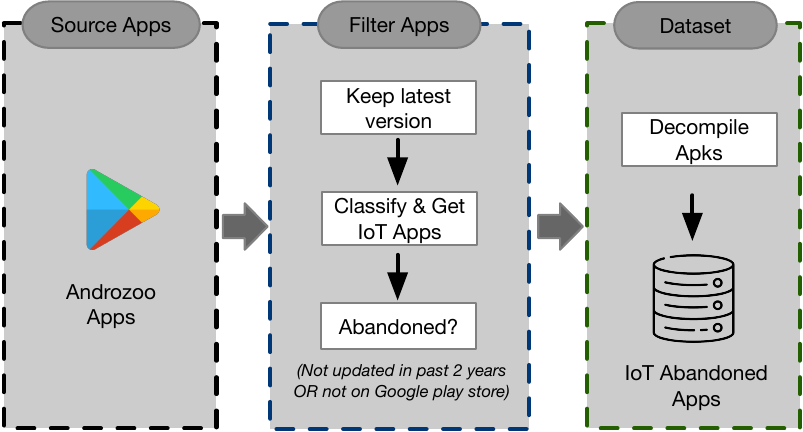}
    \caption{Dataset collection methodology.}
    \label{fig:dataset}
\end{figure}


\section{Methodology}
\begin{figure*}[t]
  \centering
  \includegraphics[width=\linewidth]{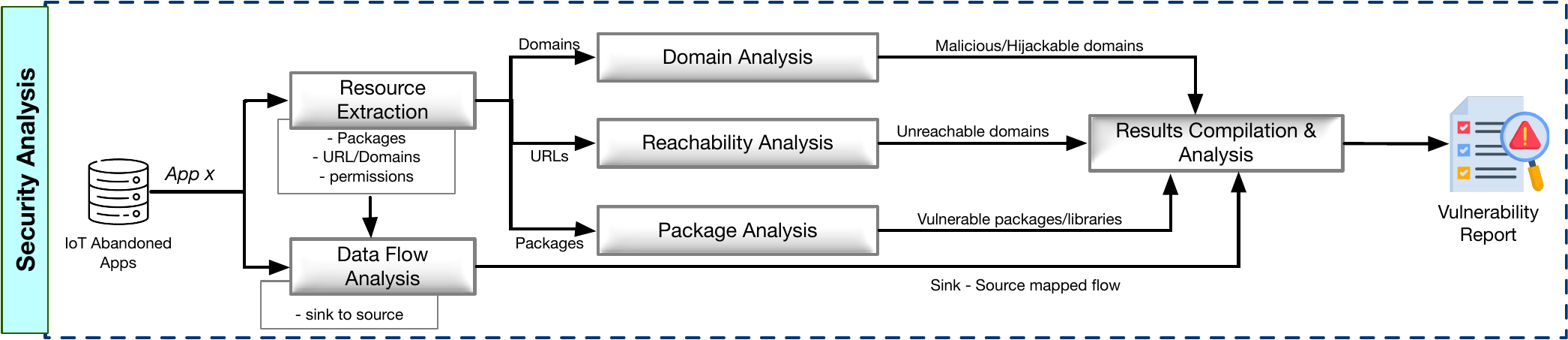}
  \caption{Analysis pipeline for security analysis of IoT abandoned apps in our dataset.}
  \label{fig:system-design}
\end{figure*}






In this section, we describe our analysis methodology for quantifying IoT abandonware risks.

\subsection{Dataset Collection}
\label{sec:dataset}

As illustrated in Figure \ref{fig:dataset}, we curated our dataset by first collecting 4,074,376 Android apps available in  AndroZoo~\cite{AndroZoo} until March 2025, which also provides metadata \cite{alecci2024androzoo} from the Google Play Store. 
Because AndroZoo periodically synchronizes with the Google Play Store, the APK snapshots do not always correspond to the latest store version. From our manual validation against 100 sampled unmaintained apps, we observed that 19\% of apps exhibited a version mismatch between the AndroZoo snapshot and the current Google Play Store listing. 
However, we characterize each app based on the AndroZoo snapshot itself rather than the current live listing since we focus on abandoned and unmaintained apps. Thus, this synchronization delay does not affect the validity of our vulnerability findings -- it only means the specific binary we analyzed for a given app may differ slightly from the live version on the Play Store.

Next, we removed duplicate versions of the same app, keeping only the latest version for each.
We then classified apps to retrieve IoT apps from the set of all Android apps collected. To do this, we evaluated several classifiers, including IoTSpotter's~\cite{jin2022understanding} BERT model, other LLMs, and zero-shot classifiers. We compared classification performance on 200 labeled apps (100 IoT and 100 Non-IoT), using application description from metadata to classify each app as IoT or Non-IoT. We selected IoTSpotter model for its high recall (97\%) compared to other models, including Phi 3.5, Gemma 3, DeBERTa for zero-shot classification, and Mistral. While prioritizing recall, manual verification of a random sample of 100 apps classified as IoT showed a 33\% false-positive rate. 
False positives include the following cases: (1) 7 apps that contain the keyword "sensor" such as "GPS", "microphone", and "camera" but are not paired with external physical devices, (2) 5 apps use the word "car", but it does not control actual cars and provides information and simulation, (3) 5 apps transferring files or mirroring screen to smart TV, (4) the remaining apps include misleading unrelated words (e.g., "sensor", "device", "bluetooth") in the description. This does not undermine our overall findings. Our vulnerability analysis is uniformly applied to each decompiled app based on its extracted binary artifacts, independent of the correctness of its IoT label. Misclassified apps mostly exhibit connectivity-related characteristics (e.g., descriptions containing "radar" or "car" keywords) and comparable library and permission profiles to true IoT apps, meaning that a 33\% false-positive rate does not introduce a structurally different distribution. 
Moreover, these misclassified apps are not expected to be more vulnerable than the broader app population. Their inclusion therefore lowers, rather than inflates, the observed vulnerability rates among apps labeled IoT---meaning our reported risk rate is a conservative (lower-bound) estimate of the true prevalence of security risk in IoT abandonware.

Using the selected BERT classifier, we filtered 95,058 apps as IoT companion apps, and checked their current status on the Play Store using Google Play Scraper~\cite{googleplayscraper}. Any app that was no longer available on the Play Store or had not received an update after March 2023 was marked as abandoned and decompiled using Jadx~\cite{jadxGithub}. In the end, we had a total of 61,500 apps in our dataset, including 54,886 apps removed from the Google Play Store and 6,614 apps unmaintained for more than 2 years.

\subsection{Analysis Pipeline Overview}

We show an overview of our analysis pipeline in Figure~\ref{fig:system-design}. For each decompiled app in our IoT abandonware dataset, we run two main analyses: Embedded Resource Analysis and Data Flow Analysis. 

In the embedded resource analysis, we extract app permissions, bundled package (library) dependencies, and static URLs/IP addresses to assess their current security posture. The permissions governing access to sensitive user data and device sensors are extracted from the \texttt{AndroidManifest.xml} file. These permissions enable the identification of the data types collected by each application. The bundled library dependencies and their version information are extracted from the \texttt{META} file. We examine them against the National Vulnerability Database (NVD)~\cite{nistSearchStatistics} to find vulnerable obsolete libraries for which CVEs are reported. Finally, we extract static URLs/IP addresses from the decompiled Java code using regular expressions. We test reachability to the associated domains/IP addresses and verify domain ownership to detect ownership changes, and finally, evaluate the URLs against known blocklists to determine if the app communicates with a known malicious endpoint.



In the Data Flow analysis, we assess potential data exposure when the data is transferred to vulnerable external endpoints. Using the static data flow analysis tool, IoTFlow~\cite{schmidt2023iotflow}, we extract data flow sinks, sources, and value sets at specific program points. From the value sets of the output, we match the value points of the known encryption algorithms and determine the encryption algorithms used in the IoT app. By combining the sink and source endpoints, network communication methods, permissions and domains, we analyze what kind of sensitive data flows into the unreliable remote endpoints, and over which network communication channel. 

Finally, we compile all analysis results and report comprehensive risks for the application.

\subsection{Embedded Resource Analysis}

\subsubsection{\textbf{Outdated Dependencies Analysis}}~\label{sec:cve-analysis}

We identify vulnerable libraries used in IoT Android applications that are reported as CVEs after abandonment. The decompiled CVE-reported libraries increase the attack surface if the IoT app vendor has not patched the library. Our algorithm is described in Appendix~\ref{sec:appendix-cve-alg}. 
From the decompiled code, we extract the library names from the subdirectory names of the \texttt{resource} directory. Then we search for the library's name in the NVD vulnerability search dataset~\cite{nistSearchStatistics}. If the search result returns too many CVEs, we use the subdirectory's name recursively to locate directly related CVEs. In this step, we filter the wrong CVE results that have a different Common Platform Enumeration (CPE) name from the library name that we found to avoid false positives. It indicates that the library name we tried to search for is included in the description of another CVE. 
When it successfully finds the CVE that matches with the CPE, we compare the version of the vulnerable library from the CPE information and the decompiled Java code. If the decompiled Java code and CPE do not have version information, we identify the possible library version via the application's last updated date from AndroZoo and the CVE published date, respectively. After matching the versions and determining whether the library used in the APK file has a reported CVE or not, we record the library name, searched CVE, version information to identify the library's vulnerability, and CVE descriptions to use for further analysis.

\subsubsection{\textbf{At-Risk Domain Analysis}}~\label{sec:hijack-domain}
IoT companion applications communicate with diverse endpoints, including cloud backends, device discovery services, firmware update servers, analytics platforms, and the IoT devices themselves. Discontinuation of these endpoints is critical because abandoned applications continue making network requests to domains that may have changed ownership, been compromised, or begun serving malicious content since the application's last update. Unlike actively maintained applications that can respond to compromised endpoints through updates, abandoned applications represent persistent attack vectors.
We perform static analysis on our dataset of decompiled APK files to extract all URL references embedded in the application code. Using the regular expression pattern matching below, we extract HTTP and HTTPS links, as well as IPv4 addresses.

$
(http| https)://([a-zA-Z0-9\textbackslash.-]+\textbar [0-9]{1,3}(\textbackslash.[0-9]{1,3}){3})(:[0-9]+)?(/[a-zA-Z0-9\textbackslash./?=\_\%:-]*)?
$


\Parhead{Unreachable Domains.} \label{sec:domain-reachability}
Abandoned applications continue attempting to contact domains long after developers cease maintenance. Understanding which of these domains remain reachable to the abandoned apps is critical for security assessment. Unreachable domains may indicate backend service shutdowns, creating domain expiration and potential takeover opportunities. Conversely, domains that remain reachable may imply either persistent cloud infrastructure or, more concerning, domains that have changed hands without the application being updated to reflect new ownership. We define reachability through DNS resolution success. In Algorithm~\ref{fqdn-algorithm}, a fully qualified domain name (FQDN) is classified as reachable if it successfully resolved to at least one valid IPv4 address record (DNS \texttt{A} record). Domains that fail to resolve, return \texttt{NXDOMAIN} responses, or have no \texttt{A} records are classified as unreachable. 

\begin{algorithm}[]
\caption{DNS Reachability Analysis}
\label{fqdn-algorithm}
\KwIn{Set of FQDNs $F = \{f_1, f_2, \ldots, f_n\}$ extracted from application}
\KwOut{Sets $R$ (reachable), $U$ (unreachable), and mapping $M: \text{FQDN} \rightarrow [\text{IP addresses}]$}

$R \gets \emptyset$; $U \gets \emptyset$; $M \gets \emptyset$\;
\ForEach{$f \in F$}{
    $records \gets \text{DNS\_QUERY}(f, \text{type='A'}, \text{recursive=True})$\;
    \eIf{$records \neq \emptyset$}{
        $R \gets R \cup \{f\}$\;
        $M[f] \gets records$\; \tcp{Store resolved IPs for later analysis}
    }{
        $U \gets U \cup \{f\}$\;
    }
}
\Return{$R, U, M$}\;
\end{algorithm}

We use recursive DNS resolution, querying Google's public DNS resolvers (8.8.8.8) to ensure consistent, geographically representative results. Recursive resolution is appropriate for this security analysis because it reflects the actual DNS behavior that abandoned applications would experience when deployed on user devices. We configure a 5-second timeout per query to distinguish genuinely unreachable domains from temporarily slow-responding nameservers.

\Parhead{Blocklisted Domains.}\label{sec:domain-blocklist}
The reachable domains identified previously continue receiving traffic from active abandoned applications. If one of the domains is taken by a malicious provider, users may be exposed to malicious content, data exfiltration, or privacy violations, and cannot be protected immediately due to the suspension of the vendor's regular patch. 
To assess the security posture of the extracted domains, we match them against curated blocklists spanning multiple threat categories as illustrated in Table~\ref{tab:domain_categories}. This analysis addresses a fundamental security question: Are abandoned applications communicating with known malicious infrastructure?

\begin{table}[!htbp]
\centering
\small
\caption{Blocklist domain categories and descriptions.}
\label{tab:domain_categories}
\begin{tabularx}{\columnwidth}{@{}lX@{}}
\hline
\textbf{Category} & \textbf{Description} \\ 
\hline
Malware & Domains serving malicious software, exploit kits, drive-by downloads, or command-and-control infrastructure \\ 
Phishing & Domains impersonating legitimate services to harvest credentials or sensitive information \\ 
Scam & Domains operating fraudulent schemes, fake stores, or deceptive services \\ 
Spyware & Domains associated with surveillance software, stalkerware, or unauthorized data collection \\ 
Tracking & Domains performing behavioral tracking, fingerprinting, or cross-site identity correlation \\ 
Ads & Advertising networks and promotional content delivery, particularly aggressive or deceptive advertising \\ 
Spam & Domains associated with unsolicited bulk messaging or email spam operations \\ 
Suspicious & Domains exhibiting anomalous behavior, newly registered with suspicious patterns, or associated with threat actors but not yet confirmed malicious \\ 
\hline
\end{tabularx}
\end{table}

To construct a robust blocklist dataset, we combine three established threat intelligence sources. 
We integrate \textit{Firebog}~\cite{firebog}, a meta-aggregator of community-curated lists evaluated for false positives and update frequency, using its advertisements, tracking, malware, phishing, and suspicious domain categories. We incorporate \textit{URL Haus}~\cite{URLhaus} by Abuse.ch for real-time intelligence on active malware distribution URLs. Finally, we include \textit{OISD}~\cite{oisd}, which uses algorithmic validation to provide broad coverage of malicious, scam, and tracking domains while minimizing false positives.

Blocklist-based threat intelligence suffers from false positive risks, particularly for high-traffic legitimate services that may appear on community-contributed lists due to overaggressive blocking policies. To address this, we implement false positive filtering using the Tranco top 1~M domains list \cite{tranco}. 

Our filtering policy operates as follows: Domains appearing in the Tranco list are excluded from malicious classification unless they match our blocklist at the subdomain level. For example, while \textit{example.com } itself might appear in Tranco's top 1M and thus be excluded, \textit{malicious-subdomain.example.com} would still be flagged if matched in our blocklists. 
This hierarchical approach reduces false positives while accounting for legitimate high-traffic domains that may contain compromised or malicious subdomains.

We match each extracted URL against the curated blocklist in a hierarchical manner. First, we check the full URL (including path and query). If no match is found, we check the full host (subdomain), and finally the base domain.
This approach reflects the varying granularity of different blocklists: Firebog blocks entire domains, OISD targets subdomains, and URL Haus specifies full URLs. The process ends at the first match, and the URL inherits the threat category of the matched entry.








\Parhead{Changed Domain Ownership.} \label{sec:domain-owner}
Domain ownership changes represent a particularly insidious security threat for abandoned applications with hard-coded network endpoints. When a domain changes hands while an application continues referencing it without updates, the new owner inherits all traffic data and trust relationships established by the previous owner. For actively maintained applications, developers can respond to ownership changes by updating endpoints, implementing certificate pinning, or issuing warnings to users. Abandoned applications lack any such remediation pathway, making them permanently vulnerable to any security consequences of ownership transfers or takeovers.

We use WHOISXML History API \cite{whoisxmlapi} to retrieve complete historical WHOIS records for domains in our dataset. The API provides timestamped WHOIS snapshots that reveal key registration details, including registrant contacts, nameservers, and registration status.
For each app, we use its last known update date (if available) or a constant date equivalent to 2 years ago from March 2025 as the baseline for representing the last point at which the application's developers exercised control over domain selection in the apps. We compare the registrant contact information, such as \texttt{name, organization, email} and \texttt{address}, across the domain records' \textit{registrantContact}, \textit{administrativeContact}, and \textit{technicalContact} data fields. We classify determinations by confidence levels based on the available identifiable information. A \textit{High} confidence level represents comparable organizational identifiers, \textit{Medium} confidence level for when only email domains or partial information is available, and \textit{Low} for when only registrant names match.

\subsubsection{\textbf{Collectible Data Type Extraction}}~\label{sec:permission-analysis} 
To reveal the data type that the abandoned app collects and possesses, we leverage the declared permissions in each application. We extract all permissions in the application's \texttt{AndroidManifest.xml} file and analyze their overall distribution across the dataset. In particular, we identify permissions on sensitive user data, such as user data storage, location, microphone, and camera. 

To explore permissions on the IoT device sensor, we automatically analyze decompiled apps to find sensor accesses (e.g., gyroscope, step counter, temperature). This helps us count how often sensitive sensors are used and identify apps that might collect more data than needed.
After analyzing the decompiled code, we search for references to Android’s sensor constants. Specifically, we scan all non-binary files and use a regular expression pattern that detects occurrences of sensor identifiers following the format \verb|Sensor.TYPE_<SENSOR_NAME>|. The pattern used was \texttt{"\textbackslash bSensor.TYPE\_[A-Z\_]+\textbackslash b"},
which matches sensor access calls (e.g.,\texttt{Sensor.\allowbreak TYPE\_GYROSCOPE}, \texttt{Sensor.\allowbreak TYPE\_ACCELEROMETER}).

We then aggregate permission occurrences and types to determine which permissions are most commonly requested and which types of data are acquired by abandoned apps. In Section~\ref{sec:data-flow}, we use output to analyze which sensitive data are most frequently retained and where the data flows to save and process user information.

\subsection{Broken Data Flow \& Cryptographic Analysis}~\label{sec:data-flow}
Using the state-of-the-art static analysis tool, IoTFlow~\cite{schmidt2023iotflow}, we analyze the reconstructed value points and their flow with two goals: (1) which sensitive data flows to the broken or at-risk external endpoints, and (2) how prevalent deprecated cryptographic primitives are across the app.

IoTFlow consists of two analysis phases: Value Set Analysis (VSA) and general Data Flow Analysis (DFA). VSA reconstructs values at specific program points, and these trigger points are sources and sinks, which include Application Programming Interface (API) keys and Universally Unique Identifiers (UUIDs). 
DFA identifies which data companion apps communicate with whom and how they share them. IoTFlow analyzes the data flow from the identified trigger points and sensitive data sources. 

For the first goal, we identify the source, sink, and network communication method that the flow uses from the DFA, and determine what kind of permissions are located at the sink or source. For example, if a flow originates from the function \texttt{getSimSerialNumber()} and is transmitted through \texttt{HttpURLConnection}, we conclude that this app transfers the sensitive user data (i.e., serial number) over Wi-Fi communication. We then match the extracted endpoints with the potentially at-risk domains list from Section~\ref{sec:hijack-domain} that indicates the possibility of data flow violations in abandoned apps.  

For the second goal, we identify each app's overall cryptographic practices in IoT abandonware. From the output of value set analysis, we scan all reconstructed values across the app and API calls corresponding to known encryption algorithms, and match these against a list of non-recommended primitives (e.g., Blowfish, DES, 3DES, GOST, and RC4). We marked the app as using a deprecated primitive if any matched value appears in the reconstructed value set, independently of a specific network call, because IoTFlow does not support linking specific value points between crypto and network calls. This yields an app-wide measure of the prevalence of deprecated cryptography, capturing latent weaknesses.

\subsection{Validation of Analysis Tools}
To assess our analysis tools, especially CVE-to-library matching and data-flow reconstruction, we manually validated them. Evaluating false negatives requires a complete ground-truth set of vulnerabilities in the analyzed IoT abandonware. Since it is infeasible to acquire an independent and comprehensive ground-truth of vulnerable apps and all identified vulnerabilities, we do not report false-negative rates. Instead, we focus on false positives. 

\subsubsection{\textbf{CVE-to-Library Matching}}
Given the scale of our dataset, we randomly sampled 100 apps from 45,248 CVE-library-matched apps.
From the decompiled APK files, we manually investigated the libraries used and their versions, focusing on \texttt{BuildConfig.java} — an auto-generated file created by Android's build system for each compiled library, typically containing build-time constants such as the library's version name and code. Since some libraries specify their version directly in the source rather than in a \texttt{BuildConfig.java} file, we additionally extracted versions from the code when this file was not present. First, we validated the analysis tool's automatic library version extraction, and the false-positive rate is 17.7\%. Second, using the correctly version-matched libraries, we manually generated the ground truth for CVE-to-library matching. This step presents a 52.5\% false-positive rate, excluding three CVEs whose applicability depends on the version of an interacting device and so cannot be confirmed from the library alone. Since IoT apps interact with external devices, some CVEs cannot be confirmed from the library alone. For example, our analysis identified that one library is affected by CVE-2023-33248, but confirming the vulnerability is conditional on the paired device: it applies only if the interacting Alexa software is running version 8960323972.

One plausible reason for the high false-positive rate is our keyword-based matching. 93 false positives have 33 unique (library name, library version) records. Investigating the pattern of false positives revealed 19 subpackage mismatches, 9 cases involving the same vendor but different services, and 5 cases that were only keyword matches. The analysis tool uses CVE descriptions and CPEs, rather than raw keyword matching alone, to avoid identifying CVEs that merely share a keyword. Even so, two further error sources remain. First, library names collide when searched against the NVD by name. Second, the NVD's automatically assigned CPEs themselves contain errors~\cite{sanguino2017software, sun2023inconsistent}. Prior work shows that similar approaches of version-based matching for Linux kernel CVEs yield a 68\% false-positive rate~\cite{helmke2023towards,fraunhoferHomeRouter}.
In addition, assessing the vulnerability of bundled libraries in IoT apps is further complicated by the fact that apps may be affected by the IoT device's vulnerable firmware or other external dependencies.
For example, the tool matched a CVE to \texttt{com.android.volley} via the keyword "Android," but \texttt{com.android.volley} has no reported CVE; the CVE (CVE-2022-2390) actually belongs to a different subpackage, \texttt{com.android.billingclient} (a subpackage mismatch). In another case, a CVE matched via the keyword "Intuit" concerned Intuit QuickBooks, which is unrelated to the Intuit library used in the Android app (same vendor, different service). Across these patterns, keyword-based matching produces false positives even when the CVE description keyword and the CPE both nominally match, since a matching CPE does not guarantee that the CVE actually corresponds to the same library or component.

\subsubsection{\textbf{Data-Flow Reconstruction}}
From 127 apps that contain reconstructed sources and sinks both, we randomly sampled 100 apps. 
We validate the reconstructed sinks, sources, and data flows from the IoTFlow output in two phases: an API call existence check, and a flow connectivity check from the source to the sink using Androguard, an independent reverse engineering tool used for cross-validation against the IoTFlow/FlowDroid/Soot-based analysis stack. For the connectivity check, we constructed a call graph and confirmed whether the source call can actually reach the method containing the sink call. The existence check exhibits 6.8\% of false positives (8 flows) due to the mismatched reconstructed sink or source. Cross-checking with Jadx decompiled code, we found that these cases include the reconstructed sinks in comments (i.e., dead code), and cases where the API call for the source was missing from the IoTFlow output even though the method existed. 13.7\% of false positives (16 flows) are caused by the absence of a call path between existing sinks and sources. Using static call graph Breadth-First Search (BFS) with a maximum of 15 hops, we found, for example, that a reported \texttt{Socket.getInputStream} $\rightarrow$ \texttt{OutputStream.write} flow was not actually connected. 
By further examining these connectivity false positives against the Jadx-decompiled code, we found that most stemmed from IoTFlow misidentifying line numbers or reporting unrelated sinks and sources as connected flows. However, 2 of the 16 flows were in fact connected (true positives) but were missed by the static call graph due to threads and callbacks; excluding these corrects the connectivity false-positive count to 14 flows.

Combined with the 8 false positives from the existence check, the data-flow reconstruction showed an overall false-positive rate of 18.8\% (22/117 flows) across 100 randomly sampled apps.

Overall, despite the observed false-positive rate, our analysis remains useful as a scalable risk candidate-generation. False positives are an inherent challenge in automated vulnerability detection based on static analysis, and the rates observed in our evaluation are consistent with the false-positive rates reported by prior approaches in related vulnerability-analysis studies~\cite{helmke2023towards,fraunhoferHomeRouter}. 

%% file: sections/eval.tex
\section{Observations} 
\label{sec:observations}

We detail our observations from our analysis by answering our four research questions \textit{RQ1 - RQ4}, investigating IoT-abandonware into two categories: removed from the Google Play Store and not updated for more than 2 years.

\subsection{RQ1: How Widespread Are Abandoned IoT App Installs and Use?
} 

\begin{table}[]
    \centering
    \caption{Availability analysis on existing IoT app dataset.}
    \label{tab:datasets}
    \scriptsize
    \resizebox{\columnwidth}{!}{%
        \begin{tabular}{@{}llllll@{}}
            \toprule
            Dataset     & Year & Available & No Update & Removed         & Total \\ \midrule
            IoTSpotter  & 2021 & 5193 &  8690 & 23900 & 37783  \\
            IoTProfiler & 2020 & 894  &  831  & 4483  & 6208  \\ \bottomrule
        \end{tabular}%
    }
\end{table}


To answer the first research question, we identified abandoned IoT companion apps in our dataset that also appear in two popular prior IoT Android app datasets, IoTSpotter~\cite{jin2022understanding} and IoTProfiler~\cite{nan2023you}, both collected in 2022. 
Since these datasets are four years old, we considered an app abandoned if it was no longer available on the Google Play Store or had not received an update in over two years, allowing us to quantify abandonment over this period.

This indicates that the IoT apps that were in service at the time the aforementioned datasets were constructed are now abandoned.
Table \ref{tab:datasets} highlights the number of abandoned apps from two datasets. 
In the IoTSpotter dataset, 63.3\% of apps are not in the Play Store and 23.0\% of apps are not updated for more than 2 years. Similarly, 85.55\% of apps in the IoTProfiler dataset are currently unavailable, and 72.2\% of the total are not in the Play Store. 

\begin{figure}[h]
  \centering
  \includegraphics[width=0.85\linewidth]{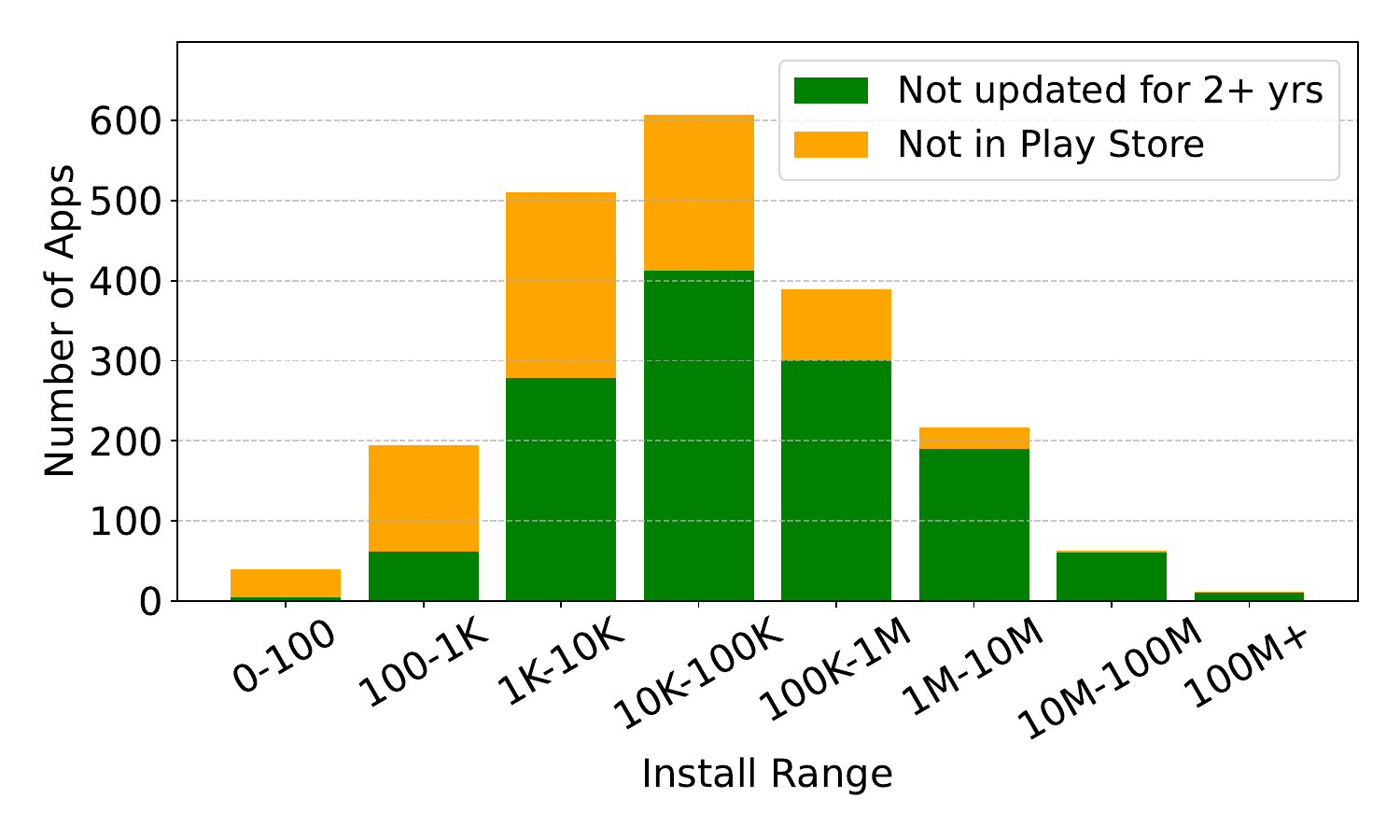}
  \caption{Download counts of abandoned IoT apps with available data in IoTSpotter and IoTProfiler.}
  \label{fig:iot-spotter-downloads}
\end{figure}


We compared the number of downloads for abandoned apps from each dataset. Figure~\ref{fig:iot-spotter-downloads} indicates that 63.4\% of the total abandoned IoT apps, 73.9\% of not updated apps, and 43.9\% of removed apps were downloaded more than 10,000 times. There are also 12 apps with more than 100 million installs from huge corporations like Huawei, Google, and Sony. 
Manual inspection of these apps suggested that they were abandoned for strategic reasons, being replaced by other apps or transformed into a different product.

\begin{findingbox}
  73.9\% of not-updated apps and 43.9\% of removed apps have been highly installed; large companies often cease regular patching or updating, even though the application is installed in more than 100 million devices. 
\end{findingbox}

\subsection{RQ2: How Prevalent Are App Artifact Vulnerabilities in IoT Abandonware?} \label{sec:rq1}

Outdated dependencies and unverified domain access in abandoned IoT apps may introduce vulnerabilities. We analyzed how such inherent risks are prevalent in abandoned IoT apps.

\subsubsection{\textbf{Outdated Dependencies}}

Obsolete libraries in the published APK files have a high potential of being vulnerable with newly reported CVEs that can be exploited by attackers as time passes after abandonment. We identify such candidate CVE-library associations in abandoned apps and assess how these vulnerabilities elevate the security risk of IoT abandonware. 

\begin{figure}[h]
  \centering
  \includegraphics[width=\linewidth]{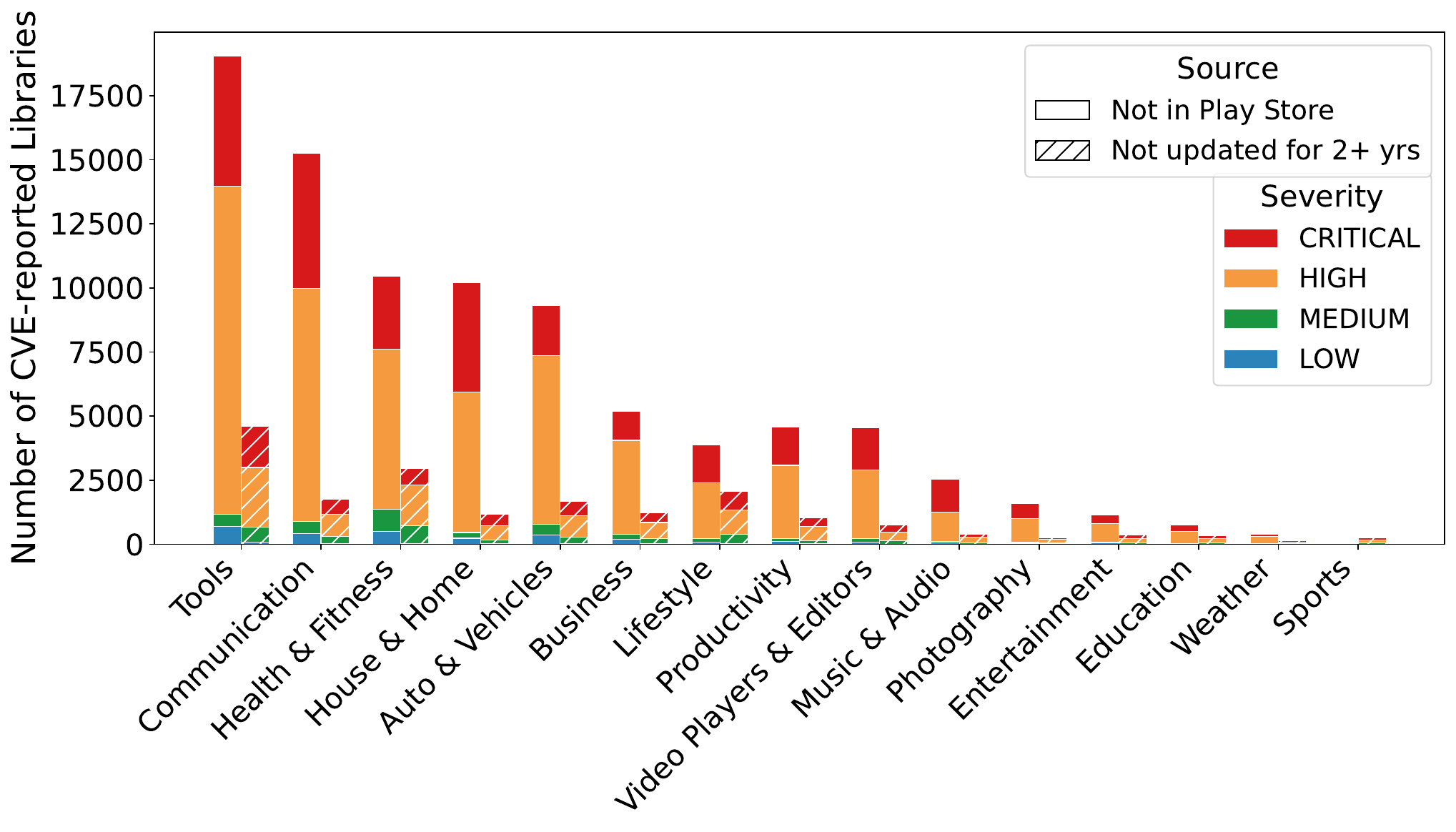}
  \caption{Results of counting CVE severity by app category for the top 15 categories.}
  \label{fig:cve-analysis}
\end{figure}


As shown in Figure~\ref{fig:cve-analysis}, 74.5\% of apps not in the Play Store and 69.9\% of unupdated apps were found to bundle libraries with potentially relevant CVEs. Among the identified CVEs, HIGH and CRITICAL severity vulnerabilities accounted for 51.0\% and 16.6\% respectively, indicating that the detected vulnerabilities were predominantly concentrated at higher severity levels.


The 10 most common libraries associated with reported CVEs in abandoned IoT apps are distributed by Android, Kotlin, Apache, OneSignal, Microsoft, Tencent, Eclipse, Tuya, Twilio, and ThreeTen. Android, Kotlin, Microsoft, Tencent, and Eclipse publish widely used libraries that are frequently included as dependencies by other libraries, which may contribute to their prevalence among the vulnerable dependencies identified in our dataset.

\begin{findingbox}
    73.6\% of IoT abandonware is highly likely bundled with CVE-reported dependencies, and removed apps possess riskier dependencies (74.5\% vs. 69.9\%); however, the bundling detection has a high false-positive rate.
\end{findingbox}

\subsubsection{\textbf{At-Risk Domains}}~\label{sec:observ-domains}
IoT apps include embedded domain names for API calls or cloud services. As the developer no longer supports or updates them, domains associated with third-party services may be withdrawn and remain as dangling resources. We tracked how many domains are vulnerable by examining domains' unreachability, change of ownership, and maliciousness. 



\Parhead{Unreachable Domains.} 
From the decompiled applications, we extracted a total of 2,799,466 FQDNs, of which 168,718 were unique. Of these unique domains, 24.2\% (40,828) were unreachable, while the remaining 75.8\% (127,890) remained active at the time of analysis. Notably, every IoT abandonware app in our dataset contained at least one unreachable FQDN, with a single app, which is no longer available on the Play Store, exhibiting as many as 4,046 unreachable domains. This underscores the extent to which abandoned applications retain unresolvable infrastructure references.

\begin{table}[]
\centering
\caption{Most common unreachable FQDNs.}
\label{tab:common-unreachable}
\resizebox{\columnwidth}{!}{%
    \begin{tabular}{@{}lrrrll@{}}
        \toprule
        \multirow{2}{*}{\textbf{FQDN}} & \multicolumn{3}{c}{\textbf{Apps}} & \multirow{2}{*}{\textbf{Purpose}} & \multirow{2}{*}{\textbf{Threat}} \\
        \cmidrule(lr){2-4}
         & \textbf{Total} & \textbf{No update} & \textbf{Removed} &  &  \\ \midrule
        ns.adobe.com                               & 28,657  & 2,697   & 25,960  & Nameserver   & No     \\
        geoip.api.p3insight.de                     & 6,076   & 123     & 5,953   & Tracking     & Maybe  \\
        awsdus.api.p3insight.de                    & 5,522   & 94      & 5,428   & Unknown      & Maybe  \\
        reports.crashlytics.com                    & 5,055   & 605     & 4,450   & Analytics    & No     \\
        settings.crashlytics.com                   & 4,744   & 496     & 4,248   & Analytics    & No     \\
        e.crashlytics.com                          & 4,741   & 496     & 4,245   & Analytics    & No     \\
        CSC3-2004-crl.verisign.com                 & 3,285   & 99      & 3,186   & Certificates & No     \\
        logo.verisign.com                          & 2,725   & 288     & 2,437   & Certificates & No     \\
        schemas.applovin.com                       & 2,348   & 80      & 2,268   & Ads          & No     \\
        ns.attribution.com                         & 2,166   & 67      & 2,099   & Nameserver   & No     \\
        casper.beckman.uiuc.edu                    & 1,920   & 190     & 1,730   & University   & No     \\
        data.flurry.com                            & 1,203   & 48      & 1,155   & Analytics    & No     \\
        csc3-2009-2-crl.verisign.com               & 1,150   & 104     & 1,046   & Certificates & No     \\
        outcome-crash-report.supersonicads.com     & 1,122   & 35      & 1,087   & Ads          & No     \\
        127.0.0.1                                  & 1,070   & 129     & 941     & Local        & No     \\
        www.google                                 & 1,027   & 114     & 913     & False +ve    & No     \\
        acs.amazonaws.com                          & 1,004   & 82      & 922     & Platform     & No     \\
        www.limbicsoftware.com                     & 959     & 81      & 878     & Unknown      & Maybe  \\
        www.dotnetdotcom.org                       & 955     & 78      & 877     & Unknown      & Maybe  \\
        www.jadynave.com                           & 955     & 78      & 877     & Unknown      & Maybe  \\
        \bottomrule
    \end{tabular}%
}
\end{table}


We analyzed some of the most common unreachable FQDNs and noted them in Table~\ref{tab:common-unreachable}, along with the purpose of the domain. By doing this, we can see that there are limitations to this methodology. Certain types of FQDNs--such as nameservers, internal endpoints, or platform-specific service URLs--may inherently appear unreachable to our probing methodology, even if they remain functional within their intended contexts. Some domains, when extracted, appear unreachable when in reality they are reachable and were extracted incorrectly because of a conflict between the TLD and the domain name (e.g., .google TLD and google.com). Analysis of the most common reachable FQDNs did not reveal any false negatives.
We assert that vulnerabilities arising from residual trust are primarily concentrated in expired domains.



\Parhead{Owner-changed Domains.} 
To assess domain ownership transitions over the preceding two years, we queried historical WHOIS records for domains in our dataset using the WHOISHistory API~\cite{whoisxmlapi}. Although our full dataset comprises 168,718 unique domains, we restricted our analysis to a representative sample of 1,000 domains due to API access costs. Of the sampled domains, only 737 had resolvable records within the WHOISHistory API's database, forming the basis of our ownership change analysis.


We subsequently observed that 66.1\% of the domains with records had all personally identifiable information (PII) redacted under European GDPR law \cite{gdprinfoGeneralData}, thereby making affirmative conclusions about domain ownership changes infeasible. 
However, for the 33.9\% domains that were not redacted, we find that 5.6\% of them had changed ownership, indicating a high security and privacy risk for 2,154 (3.5\%) applications that continue to communicate with them. Among these applications, 1,956 (3.6\%) had been removed from the Play Store, while 198 (3.0\%) were unmaintained.

\Parhead{Blocklisted Domains.}
A total of 8,566,646 static URLs were extracted, of which 11.5\% (982,647) matched blocklist entries (Table~\ref{tab:blocklist}). Of these, 4.9\% (48,451) were exact matches (\texttt{match\_level = full}), including 733 phishing, 458 scam, 72 spyware, and 70 malware URLs. Overall, 3,066 applications (5.0\%) contained at least one malicious URL. Suspicious URLs accounted for a further 302,805 matches (30.8\% of all matches), indicating potentially unsafe or untrustworthy endpoints without a specific threat classification. Advertisements and tracking URLs accounted for 65.9\% and 2.7\% of all matches, respectively, with 69.4\% of applications containing at least one advertisement or tracking URL.

Across abandonment classes, the difference was modest for malicious URLs but substantial for advertising and tracking exposure. At the application level, 5.1\% of apps (2,787/54,886) no longer available on the Google Play Store contained at least one malicious URL, compared with 4.2\% of stale applications (279/6,614) that had no updates in 2+ years. The gap was substantially larger for advertising and tracking exposure: 71.1\% of removed applications versus 55.0\% of stale applications, a 16-percentage-point difference. Despite these differences, both abandonment classes exhibit substantial exposure to potentially harmful and privacy-invasive endpoints.

\begin{findingbox}
    3.5\% of IoT abandonware has owner-changed domains, 69.4\% include at least one blocklisted domain, with removed apps accounting for the majority of apps affected by these security risks; all applications have at least one unreachable domain. 
\end{findingbox}

\begin{table}[]
\centering
\caption{Distribution of matched URLs in the blocklist, by abandonment class. \emph{Full}, \emph{Sub}, and \emph{Dom.} give the share of URLs matched at the full-URL, subdomain, and registered-domain level, respectively.}
\label{tab:blocklist}
\resizebox{\columnwidth}{!}{%
    \begin{tabular}{@{}lrrrrrrrr@{}}
        \toprule
        \multirow{2}{*}{\textbf{Category}}
        & \multicolumn{4}{c}{\textbf{Apps (No update)}} 
        & \multicolumn{4}{c}{\textbf{Apps (Removed)}} \\
        \cmidrule(lr){2-5} \cmidrule(lr){6-9}
        & \textbf{URLs} & \textbf{Full \%} & \textbf{Sub \%} & \textbf{Dom. \%}
        & \textbf{URLs} & \textbf{Full \%} & \textbf{Sub \%} & \textbf{Dom. \%} \\
        \midrule
        Ads
            & 35,344 & 4.7  & 95.1 & 0.2
            & 612,227 & 3.8  & 95.9 & 0.3 \\
        Suspicious
            & 30,228 & 5.4  & 94.6 & 0.0
            & 272,577 & 6.9  & 93.0 & 0.0 \\
        Tracking
            & 1,837  & 11.0 & 89.0 & 0.0
            & 24,636  & 5.7  & 94.3 & 0.0 \\
        Scams
            & 245    & 13.5 & 86.5 & 0.0
            & 2,512   & 16.9 & 83.1 & 0.0 \\
        Phishing
            & 131    & 61.8 & 38.2 & 0.0
            & 1,451   & 44.9 & 53.7 & 1.4 \\
        Malware
            & 90     & 5.6  & 94.4 & 0.0
            & 1,217   & 5.3  & 94.7 & 0.0 \\
        Spyware
            & 13     & 84.6 & 15.4 & 0.0
            & 122     & 50.0 & 50.0 & 0.0 \\
        Spam
            & 1      & 0.0  & 100.0 & 0.0
            & 16      & 0.0  & 100.0 & 0.0 \\
        \midrule
        \textbf{Total}
            & \textbf{67,889} & \textbf{5.3} & \textbf{94.5} & \textbf{0.1}
            & \textbf{914,758} & \textbf{4.9} & \textbf{94.9} & \textbf{0.2} \\
        \bottomrule
    \end{tabular}%
}
\end{table}

\begin{figure}[h]
  \centering
  \includegraphics[width=0.9\linewidth]{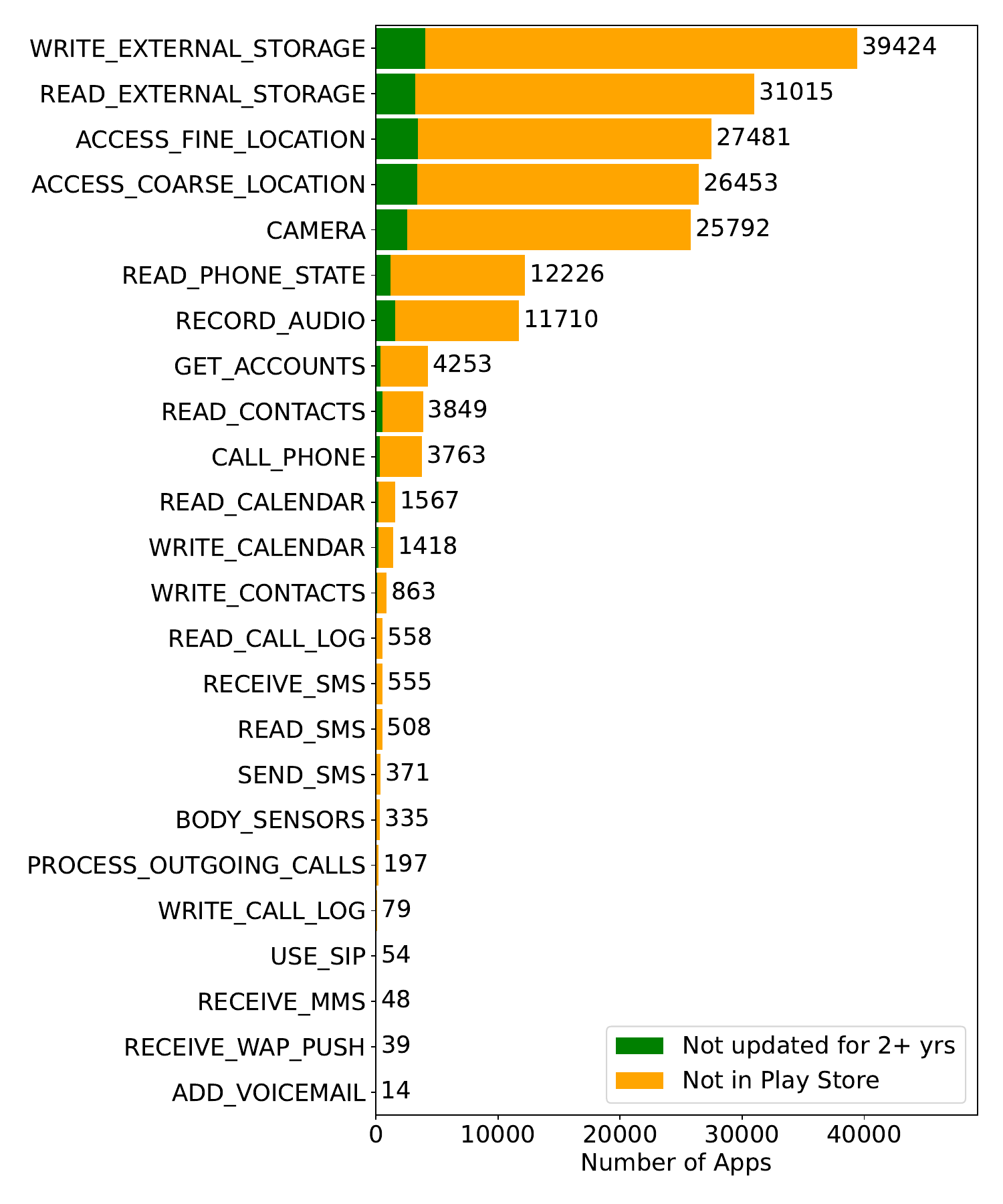}
  \caption{Distribution of dangerous permissions in 61~K abandoned IoT apps.}
  \label{fig:dangerous_permissions_chart}
\end{figure}

\begin{figure}[h]
  \centering
  \includegraphics[width=0.9\linewidth]{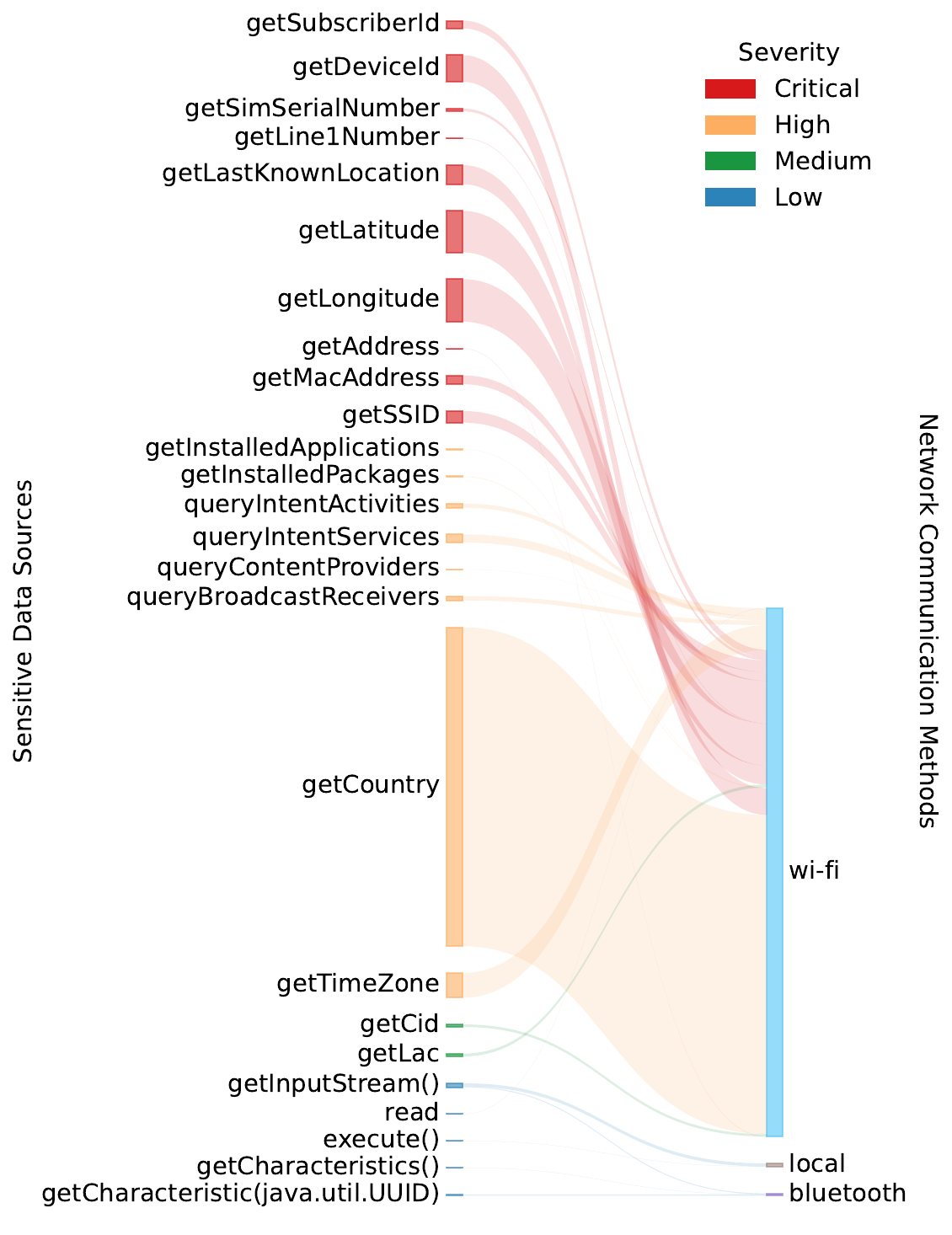}
  \caption{Result of flows from sensitive data (left) over network communication methods (right in plot).}
  \label{fig:flowanalysis}
\end{figure}

\begin{figure}[h]
  \centering
  \includegraphics[width=\linewidth]{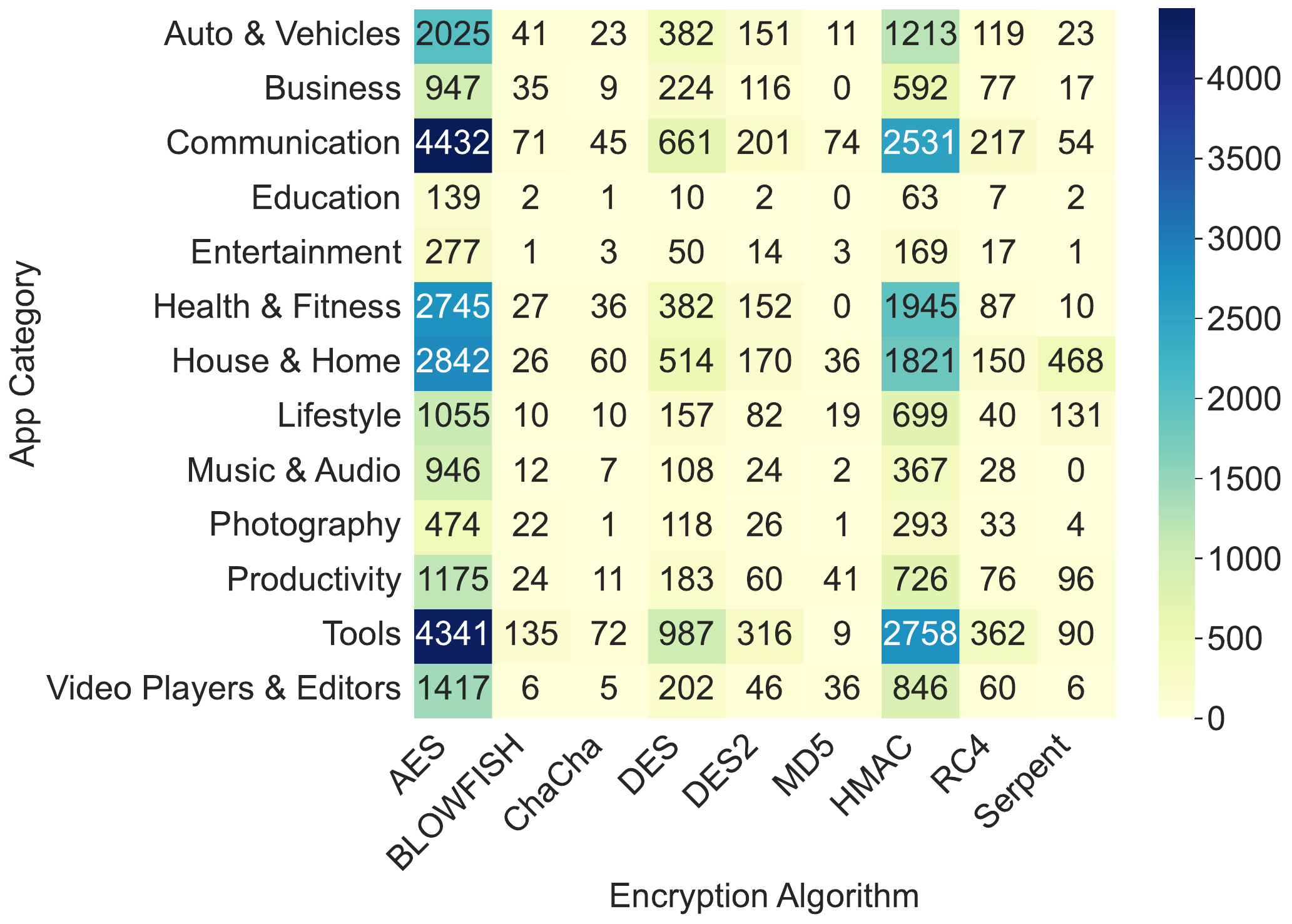}
  \caption{Dispersion of different encryption algorithms across apps not available on the Play Store.}
  \label{fig:encrypt-alg}
\end{figure}

\subsection{RQ3: What Is the Risk of Sensitive Data Exposure from Abandoned IoT Apps?} \label{sec:rq2}  

IoT companion apps differ fundamentally from general-purpose Android apps in that they require bidirectional network communication with both local sensors and remote endpoints, such as cloud APIs and external storage backends. This architectural dependency raises two critical questions: what kinds of sensitive data do abandoned IoT apps handle, and how reliable and secure is the network or communication channel on which such sensitive data is transmitted? We address both questions through a multi-faceted analysis of our IoT abandonware dataset, covering declared permissions, sensor access patterns, data flow characteristics, endpoint reachability, and cryptographic practices.

\subsubsection{\textbf{Permission and Sensor Access Patterns}}
We begin by characterizing the data acquisition surface of abandoned IoT apps through their declared permissions and hardware sensor usage. Figure~\ref{fig:dangerous_permissions_chart} presents the frequency distribution of dangerous permissions across all 61,500 apps in the IoT abandonware dataset. Permissions to access user accounts, external storage, device location, and camera feeds are predominant, collectively establishing a broad sensitive data acquisition surface even before network transmission.

Beyond declared permissions, we examine hardware sensor usage across the dataset. While sensors can provide rich contextual information, unnecessary or excessive sensor access introduces distinct security and privacy concerns. Our results show that the vast majority of apps, over 61,000 (approximately 99\%), access zero hardware sensors, indicating that most IoT companion apps rely primarily on network communication and cloud-hosted services rather than locally sampled sensor data. This finding is consistent with the functional role of many such apps, which commonly serve as remote control interfaces, monitoring dashboards, or device configuration utilities. Nevertheless, we find a small but noteworthy subset of 101 apps (92.1\% are not in the Play Store) that access more than 16 distinct sensors each. This level of sensor exposure is highly atypical for companion apps and warrants closer scrutiny from a privacy risk standpoint.

\subsubsection{\textbf{Sensitive Data Classification and Transmission Channels}}
Having characterized the data acquisition surface, we next examine what categories of sensitive data these apps collect and over which network channels that data is transmitted. We classify the data types collected by abandoned IoT apps into four sensitivity tiers~\cite{nan2023you, gdprinfoChapterController, ren2019information, girija2026digital} guided by two criteria: (i) the identifiability of the data, i.e., whether it can, alone or in combination, be used to re-identify a specific user or device
(as defined in GDPR Art.~4~\cite{gdprinfoChapterController}), and (ii) the potential
harm resulting from its exposure (e.g., financial loss, physical safety,
behavioral profiling), following the taxonomy of Nan et al.~\cite{nan2023you}.
Concretely, we define: \textit{critical} (e.g., PII and device identity data
that directly identify a user or device), \textit{high} (e.g., usage
profiling data that reveal behavioral patterns over time), \textit{medium}
(e.g., coarse-grained contextual information that is only sensitive when
combined with other signals), and \textit{low} (e.g., common API outputs and
direct user inputs with limited re-identification risk).

We manually enumerated every extracted data-collecting API method and labeled each with one of the four tiers (Table~\ref{tab:data-type-taxonomy} in Appendix~\ref{sec:appendix-taxonomy}). The resulting distribution of data flows across tiers and network channels is illustrated in Figure~\ref{fig:flowanalysis}.
98\% of observed data flows, which predominantly contain only critical and high-class sensitive user data, are transmitted over Wi-Fi connections. While this reflects the home-network-centric deployment model of most consumer IoT devices, it also concentrates risk on a single, widely targeted communication channel—one that is particularly susceptible to interception when not adequately protected.

\begin{findingbox}
    98\% of sensitive data flows in abandoned IoT apps are transmitted exclusively over Wi-Fi. As shown in Section~\ref{sec:crypto}, a large share of these apps relies on broken or legacy encryption algorithms, meaning sensitive data, including PII and device identifiers, is not only funneled through a high-risk channel but also cryptographically underprotected against interception and injection attacks.
\end{findingbox}

\subsubsection{\textbf{Data Flows to Vulnerable and Unreachable Endpoints}}

The transmission of sensitive data over Wi-Fi is particularly concerning when the destination endpoints themselves can no longer be considered trustworthy. Our static analysis reconstructed 6,376 data sinks, of which 1,089 were unique across two abandonment categories. Of these unique sinks, 42.8\% (374/874) from removed apps and 76.3\% (164/215) from unupdated apps appear in our list of unreachable, blocklisted, or owner-changed domains identified in Section~\ref{sec:observ-domains}. The sensitive data transmitted to these potentially compromised endpoints includes device and subscriber identifiers retrieved via critical API calls, as well as precise location data.

Critically, the risk is bidirectional. Of the 422 reconstructed data sources (274 unique), 31.3\% (80/256) from removed apps and 11.1\% (2/18) from undated apps are similarly unreachable, blocklisted, or under changed ownership, meaning that data injected from these vulnerable sources can propagate directly into the abandoned application. When sensitive data flows to or from such endpoints, the integrity of the communication channel can no longer be assumed. An adversary who acquires control of an expired or hijacked endpoint can trivially exfiltrate transmitted data, inject malicious payloads, or silently redirect traffic—all without any modification to the application itself. Even where destination endpoints remain nominally legitimate, sensitive data remains at risk through domain hijacking or network-level eavesdropping, particularly when transmitted over insufficiently protected channels.

\begin{findingbox}
    40.8\% of unique data sinks and 29.2\% of unique data sources (38.4\% in total) in abandoned IoT apps are associated with unreachable, blocklisted, or ownership-changed domains, to which apps continue to transmit highly sensitive data, including device identifiers, subscriber information, and precise GPS coordinates. Interestingly, unmaintained apps communicate more with at-risk sinks compared to removed apps (76.3\% vs. 42.8\%). 
\end{findingbox}

\subsubsection{\textbf{Cryptographic Practices and Residual Attack Surface}}
\label{sec:crypto}
The risks identified above may be further compounded by weak cryptographic practices within the apps themselves. 

To assess the cryptographic hygiene of abandoned IoT apps, we measured the app-wide prevalence of each encryption algorithm and retained those algorithms deployed across at least 10 categories for analysis. Figure~\ref{fig:encrypt-alg} presents the distribution of these prevalent algorithms in apps removed from the Play Store; we include HMAC in the analysis as a widely adopted message authentication mechanism. While the majority of abandoned IoT apps employ currently recommended cryptographic algorithms, a persistent subset relies on broken, obsolete, or legacy primitives—specifically DES, DES2, MD5, RC4, and Blowfish. At least one weak cryptographic algorithm is used in 8.5\% of applications, with DES exhibiting the highest deployment frequency among the non-recommended algorithms. Among unupdated apps, 8.3\% of apps use deprecated cryptographic primitives (details in Figure~\ref{fig:encrypt-alg-appendix} in Appendix~\ref{appendix:crypto}).

Taken together, these results indicate that a subset of abandoned IoT apps rely on deprecated or broken cryptographic primitives for their encryption operations. While the use of a deprecated primitive does not confirm that any particular operation within the app is exploitable, its presence indicates a latent weakness in the app's cryptographic implementation that could be leveraged by an adversary capable of reaching that code path.

\begin{findingbox}
    8.4\% of abandoned IoT apps employ non-recommended cryptography, alongside  broken or legacy primitives such as MD5, RC4, and Blowfish, indicating that weak or legacy cryptography co-occurs at the app level in a measurable portion of the IoT abandonware ecosystem.
\end{findingbox}

\subsection{RQ4: Are the Security Issues in Abandoned IoT Apps Unique to Abandonment?}
We assess whether the security issues identified in the preceding sections are a consequence of abandonment or reflect broader weaknesses endemic to the IoT app ecosystem. 
To this end, we construct a reference sample of 500 of the most-installed IoT companion apps with at least one update since March 2025, drawn from the same AndroZoo source as our abandonware dataset.
We select the most-installed apps rather than a random or representative sample of the active IoT ecosystem. These well-resourced, widely used, and actively maintained apps represent a distinct comparison against abandoned apps. 
We run our analysis pipeline on the active apps and compare the results against abandoned apps to analyze the abandonment-specific risks.

\parhead{At-Risk Domains.}
Blocklisted domain exposure is substantially more prevalent in abandoned apps, with 69.4\% (42,681/61,500) containing at least one blocklisted domain compared to 11.4\% (57/500) of active apps. Unreachable domains, however, appear broadly across both groups. We find that all abandoned apps (100\%, 61,500) contained at least one unreachable domain, and 88.8\% (444/500) of active apps also had an unreachable domain. This suggests that dead endpoint accumulation is potentially an ecosystem-wide phenomenon rather than one exclusive to abandonment. 

\parhead{Outdated and Vulnerable Packages.}
We find that 73.6\% (45,248/
61,500) of abandoned apps had at least one package dependency on a library version matching a known CVE, with 67.6\% of these CVE candidates rated critical or high severity. 
Active apps, on the other hand, showed a lower vulnerability rate of 69.8\% (349/500) than IoT abandonware, while exhibiting a comparable rate of high- and critical-severity CVEs among identified libraries (68.0\%).

\parhead{Data Flow Analysis.}
The starkest contrast between the two app groups emerges in data flow exposure and cryptographic practices. Only 0.4\% of unique data sinks in active apps were associated with vulnerable or unreachable endpoints, compared to 40.8\% in abandoned apps—a difference of approximately two orders of magnitude that represents the most pronounced abandonment-specific risk identified in our analysis. In contrast, the use of deprecated cryptographic primitives is surprisingly more prevalent in active apps (17\%, 85/500) than in abandoned apps (8.4\%, 5,174/61,500), indicating that legacy cryptographic practices are not unique to abandoned apps and remain prevalent among active apps as well.

\begin{findingbox}
    Abandoned IoT apps exhibit a significantly broader attack surface and severely broken data flows—with 40.8\% of unique data sinks associated with vulnerable or unresolvable endpoints compared to just 0.4\% in active apps—confirming that these are decay-driven risks directly attributable to abandonment. Other identified risks, including vulnerable dependencies and legacy cryptographic usage, are comparably prevalent in active apps, suggesting they are complexity-driven and endemic to the IoT app ecosystem at large; however, unlike active apps, abandoned apps offer no path to remediation as these vulnerabilities are disclosed over time.
\end{findingbox}

While this analysis characterizes security risks that accumulate once an app stops receiving updates, non-updating is not inherently negative, considering the tension between user convenience and security updates. Mathur et al.~\cite{mathur2017impact} find that users who avoid auto-updates often do so rationally due to past negative experiences with software updates rather than indifference to security. Truelove et al.~\cite{truelove2019topics} similarly find that "update" is one of the most frequent words in negative IoT app reviews, often reflecting frustration with updates that break existing functionality. 
Delayed updates are not always bad, but abandoned apps remove users from the opportunity to choose between this tension and the channel to remediation. 

\subsection{Case Studies}
\subsubsection{\textbf{Apps with High Embedded URL Counts}}
During our large-scale analysis of IoT abandonware, we encountered a few cases where certain apps contained an unusually high number of embedded URLs and domains. One app had over ten thousand unique domains, five apps had over four thousand domains, and another ten apps had over one thousand domains. Such a large volume of URLs within a single app is atypical. To better understand these anomalies, we examined two apps from the apps with the highest URL densities in our dataset.

One notable case is an AI-based driving assistant application designed to provide connected car functionalities such as hands-free calling, messaging, navigation, and media streaming. The app advertises integration with thousands of online radio stations, various navigation services (Google Maps, Waze, Yandex Maps), and its proprietary Smart Wheel IoT device. These integrations likely account for the extremely high number of URLs, as the app may embed streaming links, third-party API endpoints, and telemetry services directly within its codebase. Another example is an application that functions as a WiFi network scanner and router management tool. It includes compatibility with multiple router vendors (e.g., TP-Link, D-Link, Netgear, Huawei) and may embed vendor websites, device information databases, and numerous ad or analytics SDKs. The presence of such a large number of domains in both apps likely arises from bundled static datasets and third-party libraries rather than direct functional necessity, raising potential security and privacy concerns related to excessive external dependencies and domain reuse risks. 


\subsubsection{\textbf{Vulnerability Assessment of Most Installed Apps}}~\label{sec:vul-assessment}
We synthesized the results of all components of our analysis pipeline to identify the risk of the top 1,000 most-installed abandoned IoT apps. Our risk assessment methodology incorporated three primary attributes to evaluate severity within the IoT-abandonware dataset: (1) the number of dependent libraries with certainly reported CVEs, (2) inconsistent and risky domains that serve as information flow sinks, and (3) insecure data flow to unintended endpoints.

We quantified the IoT app risk of each attribute; 
confirmed high or critical CVE-reported libraries, at-risk domains (i.e., unreachable, owner-changed, and malicious domains (malware, phishing, spyware)), broken data flows with risky domains 
as a sink or source, and flagged applications employing deprecated cryptographic protocols for endpoint communication. Then, we aggregated these attributes and identified highly vulnerable apps that send sensitive user data to at-risk domains with insecure encryption algorithms. 

\Parhead{Risk characterization of popular applications.} Risk analysis of the top 1,000 most installed apps, regardless of two abandonment categories, revealed several distinctive patterns. We counted the apps that contain CVE-reported libraries (i.e., the extracted library version or published date matches the CVE-reported version or date), and 39.6\% of the top 1,000 installed apps have at least one and at most three CRITICAL CVEs, 62\% have HIGH CVEs. These apps contain an average of 148 hard-coded URLs within their compiled binaries. Notably, all top 1,000 apps include unreachable or blocklisted domains, though not necessarily malicious ones. 
0.9\% of the top 1,000 applications contain more than 100 hard-coded unreachable endpoints, 4.6\% contain malicious domains, and 3.8\% exhibit owner-changed domains. 
In particular, 5 apps, installed approximately 100,000 to 5,000,000 times, possess every type of risky domain, allowing attackers to easily initiate attacks via multiple routes by exploiting them. 

To detect vulnerable data flow to external endpoints, we 
verified that the restored sinks and sources from the data flow analysis are unreachable, blocklisted, or owner-changed domains. Among 1,000 applications, 30 apps send data to broken external endpoints, especially three apps communicate with malicious blocklisted domains. From the analysis result of encryption algorithm usage, 10\% of apps use deprecated cryptographic algorithms, and three of them present the co-occurrence of communication with risky endpoints and deprecated cryptographic usage. Given the permissions granted to the 127 apps that are the union of possessing the attack surfaces (i.e., broken data flows or deprecated cryptographic practices), the user's location, camera, audio, contacts, calendar, and external storage data can be leaked by exploiting broken backend communication.

In our entire IoT abandonware dataset, 73.62\% of the total apps contain at least one severely deprecated embedded resource -- malicious blocklisted domains, owner-changed domains, and high or critical CVE-reported libraries. From the reconstructed data flow, 1,444 apps send data to risky domains (i.e., unreachable, owner-changed, and any blocklisted domains), and 5,174 apps of the total use a non-recommended encryption algorithm. The apps with at least one identified vulnerability, as discussed above, exhibit characteristics that potentially provide a latent attack starting point.

Notably, the top 1,000 most-installed abandoned IoT apps have slightly more vulnerable 
embedded resources then the overall dataset and 0.7\% more communication with the broken backend (3\% for the top 1,000 vs. 2.3\% overall). Although widely installed apps are generally expected to be well-patched and at lower risk after abandonment, they contain more vulnerabilities in reality.

\begin{findingbox}
    Top 1,000 most-installed abandoned apps exhibit more vulnerabilities than the overall dataset (10\% weak cryptography usage and 3\% broken backend communication), contradicting the common assumption that popularity correlates with better security maintenance.
\end{findingbox}

%% file: sections/limitations.tex
\section{Limitations}
~\label{sec:limitations}

\Parhead{BERT Model Classification.}
The IoT app classification step depends on the accuracy of the IoTSpotter BERT model. Misclassification can occur when an app’s description is vague, misleading, or incomplete. Some IoT-related apps might not explicitly describe their functionality, leading to false negatives.

\Parhead{Overlooking Dynamic Behavior.}
Static code and manifest analysis may miss runtime behaviors, which can introduce new URLs or communication patterns not visible without dynamic analysis.

\Parhead{DNS-Based Reachability.}
The above analysis highlights many limitations of only checking the reachability of FQDNs using active DNS queries. This analysis could benefit from a multi-pronged detection technique using both active and passive DNS scanning\cite{Lever_Walls_Nadji_Dagon_McDaniel_Antonakakis_2016} as well as WHOIS record verification.


\Parhead{Permission Analysis.}
Permissions analysis is based on what is requested in the app’s code, but this does not necessarily reflect runtime behavior. Some permissions may not be used by the app at all, making the analysis of actual permission abuse more challenging. It is difficult to assess the full extent of misuse of permissions without knowing the full context of the app's operation, including how third-party services or backend systems interact with the app.

\Parhead{Keyword-based CVE-to-Library Matching.}
Our keyword- and version-based matching approach is prone to false positives, a challenge also observed in prior vulnerability and version identification studies~\cite{chen2025vulnerability, sun2023inconsistent, sun2023aspect}. To mitigate this limitation, we plan to incorporate LLM-based semantic modeling that accounts for the complexity of IoT ecosystems as future work.


%% file: sections/relatedwork.tex
\section{Related Work}

\Parhead{IoT Security and Privacy.} 
IoT ecosystems remain insecure across their perception, network, and application layers, with common attack vectors including unprotected communication channels and poor lifecycle management. Prior works~\cite{riahi2013systemic, neshenko2019demystifying, hwang2015iot, tawalbeh2020iot, zheng2018user, 205156, sasaki2025am_i_infected, junior2019beware, mahmoud2015internet} corroborate this at scale, finding insecure firmware, exposed services, hard-coded credentials, and weak update regimes to be pervasive across device classes. These works motivated mitigation approaches from technical isolation~\cite{10646731,184449, 8835392, deep2019iot_survey, hassan2019current} to regulatory controls~\cite{Ohm_Kim, Kohno_Acar_Loh}. However, these approaches treat the IoT vendor as a fixed accountable entity where security and privacy are analyzed as properties of an ongoing device-vendor relationship, and do not examine the risks accumulated when that relationship ends.


\Parhead{IoT Abandonware.} 
Recently, a small but growing literature has examined post-support risks, focusing on the hardware/device~\cite{Wang_Jiang_Chang_Zhou_Hou_Luo_Wu_Ren_2021, Lechelt_Gorkovenko_Soares_Speed_Thorp_Stead_2020} and network components~\cite{yaben2024towards, soos2019security}, while largely overlooking companion apps, where embedded resources (endpoints, package dependencies, etc.) pose residual risks after abandonment.

\Parhead{Residual Trust after Service Termination.}
Lever et al.~\cite{Lever_Walls_Nadji_Dagon_McDaniel_Antonakakis_2016} first demonstrated at scale that expired domains retain residual DNS trust that can be hijacked for phishing, data exfiltration, or service takeover, and proposed an algorithm to broadly detect such domains. Subsequent work extended this finding to dangling domains and URLs more generally~\cite{liu2016all, zhang2023detecting, zhang2024rethinking}, and Pariwono et al.~\cite{pariwono2018don} applied Lever et al.'s method to the mobile app ecosystem, identifying 3,628 abandoned internet resources still referenced by 7,331 in-service apps. However, this line of work examined resources left by active applications, leaving open the inverse case: applications that are themselves abandoned yet remain in active use by users, along with the residual resources and permissions they retain.


\Parhead{Sensor Access as Attack Surfaces.} 
Seemingly benign sensors (accelerometer, gyroscope, microphone, etc.) have been shown to leak sensitive information through side channels, enabling inference of speech, keystrokes, PINs, and other private inputs~\cite{Mahdad2022EarSpy, abbas2025active, Javed2020AlphaLogger, Spreitzer2014PIN}. However, these works assume an actively exploited channel and do not consider sensor access that persists in abandoned apps, regardless of whether the data's destination remains reliable.

\Parhead{Bridging the Gap.} Taken together, the literature leaves a consistent blind spot: studies of abandonware concentrate on device and network components rather than companion apps; studies of residual trust examine resources abandoned by active apps rather than abandoned apps that remain in use; and studies of sensor side channels assume an active attacker rather than a passive erosion of trust in an unmaintained app's data destinations. Our measurement study is positioned at exactly that intersection.

%% file: sections/conclusion.tex
\section{Conclusion}

We conduct the first large-scale empirical measurement study of abandoned IoT companion applications. 
We collected a dataset of 61,500 abandoned Android IoT apps (i.e., not updated for two years or no longer in service as of March 2025).
We extract latent and embedded resources (dependencies, domains, and permissions) from the apps and analyze the data flows to assess both internal app artifacts and external ecosystem risk. Our large-scale study finds that 73.6\% of apps include libraries with known CVEs, although this figure is subject to a high false-positive rate. Furthermore, every IoT app in our dataset includes at least one deprecated domain, and 38.4\% of the data flow communicates with a broken backend. Overall, our compiled analysis indicates that 73.6\% of IoT‑abandonware apps possess potential security vulnerabilities.